\documentclass[twocolumn,pra,aps,a4paper,showpacs,floatfix]{revtex4}
\usepackage{amsmath}
\usepackage{amsfonts}
\usepackage{eufrak}
\usepackage{amssymb}
\usepackage{bbold}
\usepackage{epsfig}
\usepackage{graphicx}
\usepackage{color}

\begin{document}

\title{
Fast dynamics and spectral properties of a multilongitudinal-mode semiconductor
laser:  evolution of an ensemble of driven, globally coupled nonlinear modes
}
\author{G.P. Puccioni$^{\rm 1}$ and G.L. Lippi$^{\rm
2,3,4}$\footnote{Corresponding author}}
\affiliation{$^{\rm 1}$ \mbox{Istituto dei Sistemi Complessi, CNR, Via Madonna
del Piano 10, I-50019 Sesto Fiorentino, Italy}\\
$^{\rm 2}$ \mbox{Institut Non Lin\'eaire de Nice, Universit\'e de Nice Sophia
Antipolis}\\
$^{\rm 3}$ \mbox{CNRS, UMR 7335,
1361 Route des Lucioles, F-06560 Valbonne, France}\\
$^{\rm 4}$ email:  Gian-Luca.Lippi@inln.cnrs.fr}

\date{\today}


\begin{abstract}
We analyze the fast transient dynamics of a multi-longitudinal mode
semiconductor laser on the basis of a model with intensity coupling.  The
dynamics, coupled to the constraints of the system and the below-threshold
initial conditions, imposes a faster growth of the side modes in the initial
stages of the transient, thereby leading the laser through a sequence of
states where the modal intensity distribution dramatically differs from the
asymptotic one.  A detailed analysis of the below-threshold, deterministic
dynamical evolution allows us to explain the modal dynamics in the strongly
coupled regime where the total intensity peak and relaxation oscillations take
place, thus providing an explanation for the modal dynamics observed in the
slow, hidden evolution towards the asymptotic state (cf. Phys. Rev. A {\bf
85}, 043823 (2012)).  The dynamics of this system can be interpreted as the
transient response of a driven, globally coupled ensemble of nonlinear modes
evolving towards an equilibrium state.  Since the qualitative dynamics do not
depend on the details of the interaction but only on the structure of the
coupling, our results hold for a whole class of globally, bilinearly coupled
oscillators.  (All figures in color online).  \end{abstract}

\maketitle 

\section{Introduction}\label{intro}

The transient turn-on dynamics of lasers is a topic which has received a
considerable amount of attention over the years (cf.,
e.g.,~\cite{Tang1963,Fleck1964,Lau1983,Lau1984,Baer1986,Bracikowski1991,Mecozzi1991,Stamatescu1997}),
not only for the fundamental understanding of its evolution but also, and
especially, for the importance that it represents in data encoding in
telecommunications with directly-modulated semiconductor lasers.  While
Multiple Quantum Well~\cite{Yacomotti2004,Furfaro2004} (or even Quantum
Dot~\cite{Tanguy2006}) lasers have been proven to possess a peculiar modal
dynamics --  modelled with the help of mutual nonlinear coupling and
noise~\cite{Yacomotti2004,Ahmed2002,Ahmed2003}, with a Complex Ginzburg-Landau
approach~\cite{Serrat2006} or based on multiscale analysis~\cite{Gil2011}
where the resulting model provides proof for a true phase
instability~\cite{Gil} --, inexpensive, edge-emitting devices show a
dynamics~\cite{Byrne} which is more appropriately characterized by
cooperation, rather than competition~\cite{Haken1983}.

For these latter devices, we have recently shown~\cite{slowmodes} that a
realistic model for a multimode semiconductor laser, with experimentally
matched parameters~\cite{Byrne}, predicts a slow, hidden dynamical
mode-evolution governed by a master mode.  An overall agreement exists between
our recent predictions and a wealth of experimental data
(e.g.~\cite{Lau1983,Lau1984}) since it has been time and again shown that a
gradual line narrowing exists in the progress of the multimode laser transient
(cf. also previous theoretical calculations,
e.g.,~\cite{Marcuse1983,Osinski1985,Clarici2007}).  However, no specific
experiments seem to have been carried out to quantify the physical features of
the multimode dynamics in edge-emitting semiconductor lasers (beyond its more
technical aspects), and the physical origin of our predictions concerning the
slow components of the dynamics remain so far unexplained. 

In this paper, we analyze the deterministic dynamics of the multimode laser
transient and determine the modal intensity distribution at the end of the
fast transient starting from a below-threshold condition and leading into the
slow dynamics~\cite{slowmodes} with the characteristic oscillations of a
Class-B laser~\cite{josab2}.  The {\bf apparent} {\it equilibrium solution},
taking place at the end of the fast transient, in reality amounts to an
inherent {\it out-of-equilibrium one}, when analyzed in terms of modal
intensity distribution.  Our interest for answering the questions related to
this non-equilibrium intensity distribution is of a fundamental nature and
extends, beyond strongly multimode semiconductor lasers, to all those other
types of laser -- e.g. fiber lasers~\cite{Hunkemeier2000} or solid state
lasers~\cite{Pan1992} -- which are characterized by the simultaneous
operation, at least in transient, of a large number of longitudinal modes. 

From the point of view of dynamical systems, our model can be viewed as an
ensemble of globally coupled modes possessing a common, stable attractor. 
Globally
coupled systems have been, and are still, a topic receiving a large amount of
attention given their potential for application to various fields (cf.,
e.g.,~\cite{Kaneko90,Tsang91,Parravano98,Takeuchi11}).  Multimode lasers and
laser arrays have been recognized very early on as interesting physical
examples of globally coupled systems (cf.,
e.g.,~\cite{Baer86,Wiesenfeld90,Kourtchatov95,Kozyreff00}).  At variance with
most other investigations, we do not study synchronization or the appearance
of collective states, but the transient evolution towards a fixed point
(attractor) by a collection of modes with unequal degrees of coupling to a
global energy source.  However, transient dynamics can become part of
self-sustained dynamics if the system parameters are modulated on a time scale
comparable to (at least one) of its internal constants.  

The results are non-trivial and show the existence of
multiple time scales and of collective dynamics which contribute to a fast
evolution (studied in this paper), which precedes the slow, hidden dynamics
already presented elsewhere~\cite{slowmodes}.
The discussion is entirely cast in terms of laser physics, but, making
abstraction from the direct physical meaning of the variables -- i.e.,
identifying the carrier number as a global coupling field and the laser modes
as oscillators nonlinearly coupled to that (mean) field --, the results can be
transposed to a generic dynamical system consisting of a large number of
oscillators each individually coupled to a global field.  

The model is briefly
discussed in Section~\ref{modelsect} and the transient evolution is analyzed
in Section~\ref{analysis}, which is divided into three subsections.  The first
two provide a detailed analysis of the below-threshold, deterministic dynamics
for the population -- where an analytical solution can be found for the
carrier density (Section~\ref{carrier-density}) -- and for the modal intensity
distribution -- describable in terms of approximate iterative analytical
solutions (Section~\ref{decoupled-modes}).  This detailed analysis paves the
way for the central point of the paper:  the analysis of the strongly coupled
(multimode) transient in the oscillatory regime
(Section~\ref{strong-coupling}).  This regime, which displays the
characteristic damped oscillations both in the carrier number and in the total
intensity, connects the below-threshold dynamics with the slow modal one
previously reported~\cite{slowmodes}.  In particular, its final state --
corresponding to the disappearance of  oscillations in the carrier density and
in the total laser intensity -- is responsible for the ensuing, hidden modal
dynamics~\cite{slowmodes}.  The paper concludes with the analysis of the
characteristics of the frequency spectrum emitted by the laser during the fast
transient (Section~\ref{spectra}).  Comments, a summary and conclusions are
offered in Section~\ref{conclusions}.

\section{Model}\label{modelsect}

We resort to a standard model, where the physical constants are determined by
comparing its predictions to experimental results~\cite{Byrne}, for the study
of the fast transient response of a semiconductor laser to the sudden
switch-on of its pump (control parameter).  The details, numerical values and
labeling choices for the simulations we perform have been published
in~\cite{slowmodes}. 

The physical description is based on an ensemble of $M$ lasing modes,
intensity-coupled to the carrier number $N$:
\begin{subequations}\label{model}
\begin{align}
\label{modela}
\frac{dI_j(t)}{dt} & =  [\Gamma G_j(N) - \frac{1}{\tau_p}] I_j + \beta_j
B N (N+P_0) ,\\
\label{modelb}
\frac{dN(t)}{dt} & =  \frac{J}{q} - R(N) - \sum_j \Gamma G_j(N) I_j ,
\end{align}
\end{subequations}
where $I_j(t)$ is the intensity of each longitudinal mode of the
electromagnetic (e.m.) field $(1\le j \le M)$, $N(t)$ is the number of
carriers as a function of time, $\Gamma$ is the optical confinement factor,
$G_j(N)$ is the optical gain for the $j$-th lasing mode, $\tau_p$ is the
photon lifetime in the cavity, $\beta_j$ is the fraction of spontaneous
emission coupled in the $j$-th lasing mode,  $B$ is the band-to-band
recombination constant, $P_0$ is the intrinsic hole number in the absence of
injected current, $J$ is the current injected into the active region, $q$ is
the electron charge, and $R(N)$ is the incoherent recombination term
(including radiative and nonradiative recombination) which represents the
global loss terms for the carrier number (i.e., the population inversion).  

The bracket multiplying $I_j$ in eq.~(\ref{modela}) represents the global gain
(effective modal gain minus losses) for the $j$-th mode, while $\beta_j [B N
(N+P_0)]$
represents the effective mean contribution of the spontaneous emission to each
mode.  $\frac{J}{q}$ represents the normalized current (i.e.  carrier
number injected into the junction -- energy provided to the laser) in
eq.~(\ref{modelb}), and the last term accounts for the global carrier number
depletion due to the stimulated emission into all lasing modes.

The optical gain function, $G_j(N)$, contains information about the carrier
number, $N_0$, necessary to achieve transparency  (i.e., no absorption:
$N=N_0$):
\begin{eqnarray}
\nonumber
G_j(N) & = & g_p(N-N_0)(1-\epsilon I_t) \\
\label{coeff2}
& & \qquad \left[ 1-2 \left(
\frac{\lambda_{j}-\lambda_p}{\Delta\lambda_G} \right)^2\right] \, .
\end{eqnarray}

The parameters are: $g_p$ differential gain, $\epsilon$ gain compression
factor (multiplying the total intensity $I_t \equiv \sum_j I_j$), the
individual mode wavelength
$\lambda_j$, the wavelength at the peak of the gain curve $\lambda_p$, and the 
Full-Width-at-Half-Maximum (FWHM) of the gain curve itself, $\Delta \lambda_G$. 
See~\cite{slowmodes} for
further details.

As in~\cite{slowmodes}, we have allowed for $M=113$ modes to take part in the
dynamics in order to include in the simulations modes extending over
approximately about 1.5 times the FWHM of the gain curve (cf.~\cite{slowmodes}
for additional details).  This choice allows us to give a good description of
the transient regime -- the focus of this paper -- since the comparatively
large contribution of the spontaneous emission to the side modes renders them
an important element in the initial phases of the transient and plays a
crucial role in the determination of the non-equilibrium distribution at the
end of the {\it fast transient}, thereby considerably enlarging the transient
power spectrum.

\begin{figure}[ht!]
\includegraphics[width=0.9\linewidth,clip=true]{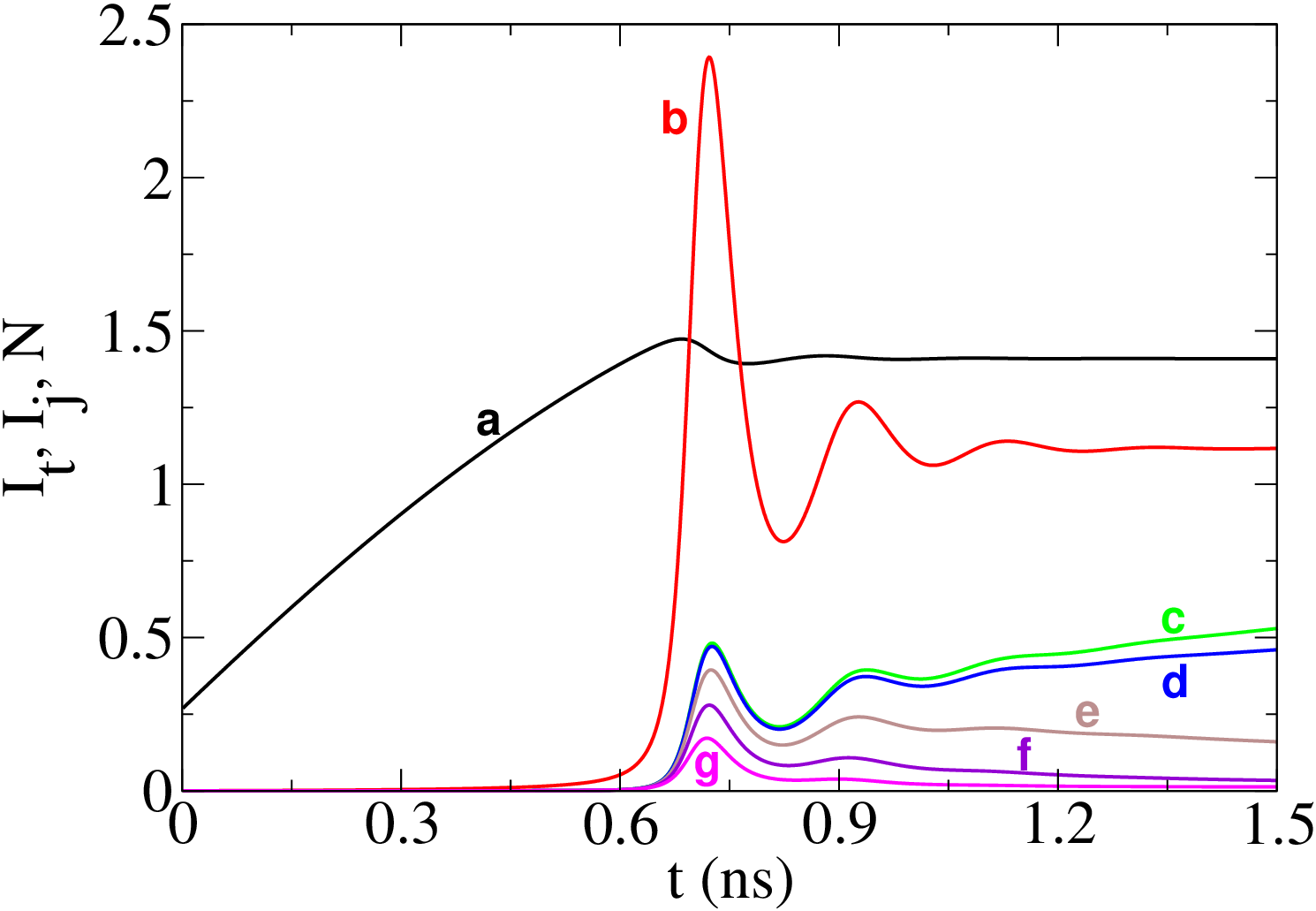}
\caption{
Time evolution of the total laser intensity, $I_t$ (curve (b)), following a
sudden switch-on of the injected current (pump rate) in form of a Heaviside
function centered at $t = 0 ns$.  Initial current value:  $J_i (t \leq 0^-) =
1.0 mA$, final current value $J_f (t \geq 0^+) = 37.5 mA$.  The threshold
current for this laser is $J_{th} = 17.5 mA$.  Model with $ M = 113$ modes.
All other parameters as in~\cite{slowmodes}.  All parameter values are kept
constant throughout the paper.  With these
values of $J_i$ and $J_f$ the intensity peak is reached at $t_p \approx 0.72
ns$, while the steady state is attained at $\overline{t} \simeq 1.5 ns$.  Here
and in all subsequent figures {\it (unless otherwise specifically noted in the
figure caption)} the total intensity $I_t$ is divided by $2.5 \times 10^5$,
the individual mode intensities by $1 \times 10^5$ and the carrier number $N$
is divided by $10^8$.  Curve (a) shows the carrier number as a function of
time.  The other curves show the individual mode intensities for the central
mode ($j = 57$, (c)), and side modes close to line center (blue side):  $j =
56$ (d), $j=54$ (e), $j = 52$ (f) and $j = 50$ (g).  The symmetric modes,
relative to the central one, are not shown but are identical.  Notice that in
the first phases of the transient (peak) curves (c) and (d) are superposed on
this scale:  they separate only at the second oscillation.  }
\label{time-trace} \end{figure}

We numerically integrate the model equations, eqs.~(\ref{model}), in response
to a sudden switch of the injected current $J$ (in form of a Heaviside
function), obtaining for the total laser intensity $I_t$ the response shown in
Fig.~\ref{time-trace} (curve (b)). 

We notice the standard delay at turn-on for $I_t$, relative to the application
of the pump-switch at $t=0$, followed by the usual relaxation oscillations
with rapid convergence towards steady state at $t = \overline{t} \approx 1.5
ns$.  The transient, is however, composed of the sum of all individual
transients and its apparent {\it usual} behaviour -- i.e., that of a
single-longitudinal mode laser~\cite{Dokhane2002} -- is far from trivial.
Fig.~\ref{time-trace} also shows a selection of lasing modes.  The top two
curves display the total intensity $I_t$ (b) and the carrier number $N$ (a),
while the lower curves represent the temporal evolution of selected modes near
line center (cf. figure caption).  In the fast transient ($t \le
\overline{t}$), the individual modes all follow the same oscillatory
behaviour, contributing to the global oscillation for $I_t$ (and $N$).
However, we remark that $\frac{I_j(\overline{t})}{I_t(\overline{t})} \ne
\frac{I_j(t \rightarrow \infty)}{I_t(t \rightarrow \infty)} \equiv
\frac{\overline{I}_j}{\overline{I}_t}$ for all modes (even though
$I_t(\overline{t})= \overline{I}_t = I_t(t \rightarrow \infty)$), which
represents the starting point for the slow dynamics~\cite{slowmodes}.

\section{Analysis}\label{analysis}

The fast transient, leading to the intensity peak and the usual damped
oscillations, can be divided into two parts.  A first one, where the carrier
number $N$ can be decoupled from the modal intensities $I_j$'s, and a second
one where -- due to the strong coupling -- the fully nonlinear system must be
retained.  

In the decoupled regime, the carrier number plays the role of an energy
reservoir, virtually unaffected by the presence of the laser modes (too weak
to have an impact on the carriers).  Thus, it is possible to find an
analytical solution for its evolution -- starting from an initial condition --
towards a transient final state defined by the breakdown of this approximation
(Section~\ref{carrier-density}). In this regime, the modal intensities are
decoupled from one another
(Section~\ref{decoupled-modes}) and evolve under the action of the
(time-dependent) energy reservoir, $N$, whose behaviour has been analytically
obtained with good accuracy in Section~\ref{carrier-density}.  This
approximate but accurate treatment allows us to define a ``threshold" for the
multimode laser (including spontaneous emission) in analogy with a basic
single mode laser model, where spontaneous emission is traditionally
neglected~\cite{Narducci}.  The strongly coupled regime is analyzed in detail
in Section~\ref{strong-coupling} on the basis of the knowledge acquired in
Sections~\ref{carrier-density} and~\ref{decoupled-modes} and of {\it ad hoc}
considerations.

The analysis of the fast transient is best conducted by choosing an initial
state with the laser (almost) entirely off.  A small amount of prebias ($J_i =
1.0 mA$, well below the lasing threshold~\cite{initialbias} which is placed
around $J_{th} \approx 17.5 mA$) is useful since it provides an average
contribution of the spontaneous emission to each mode, without having to wait
for the carrier number to grow.  This choice is not restrictive and does not
limit the validity of our results.  Our
analysis holds any time the laser starts from an injection current value below
threshold, which corresponds to an initial energy repartition among modes
which strongly differs from the final one, above threshold.  If instead the
laser is already
pumped above threshold and we suddenly change its pump to another value above
threshold~\cite{dokhane2001,dokhane2004}, the ensuing dynamics will reflect
the slow modal intensity redistribution examined in~\cite{slowmodes}.

\subsection{Decoupled carrier dynamics}\label{carrier-density}

A good understanding of the initial phases of the laser turn-on requires a
description, even though approximate, of its evolution away from noise.  A
good deal of work has been dedicated to an analytical or semi-analytical
description of this
question~\cite{Petermann1978,Agarwal1993,Zhang2007,Ab-Rahman2010a,Ab-Rahman2010b,Ab-Rahman2011,Hisham2012,Sokolovskii2012}.
In the following, we briefly outline the (standard) derivation of an
approximate analytical solution for ease of comparison with the full,
numerical integration of the model.  The quantitative comparison will show
that the analytical approximation provides excellent results even in the
initial stages of the transient when the e.m. field intensity reaches
relatively high values, i.e., exactly where one would expect the approximate
solution to already fail.

The
contribution of the spontaneous emission (intrinsic noise) to each mode is
quite small.  As a first approximation, we can therefore consider that the
initial phases of the transient are well described by the set of
eqs.~(\ref{model}) without the coupling term between individual modal
intensities $I_j$ and carrier number $N$ in eq.~(\ref{modelb}).  In this
approximation, we start by integrating the carrier number, eq.~(\ref{modelb}),
by variable separation:

\begin{eqnarray}
\label{genintegral}
\int_{N_i}^{\mathcal{N}} \frac{dN^{\prime}}{p(N^{\prime})} & = & \int_0^t d
t^{\prime} \, ,
\end{eqnarray}
where
\begin{eqnarray}
\label{poly}
p(N) & = & a N^3 + b N^2 + c N + d \, , \\
\label{coeffa}
a & = & - C \, , \\
\label{coeffb}
b & = & - (B + 2 C P_0) \, , \\
\label{coeffc}
c & = & - (A + B P_0 + C P_0^2) \, , \\
\label{coeffd}
d & = & \frac{J}{q} \, ,
\end{eqnarray}
with the constants given in~\cite{slowmodes}.

Formal integration of the left-hand-side (l.h.s.) of eq.~(\ref{genintegral})
gives~\cite{intwithmathematica}
\begin{eqnarray}
\label{formalsol}
\int_{N_i}^{\mathcal{N}} \frac{dN^{\prime}}{p(N^{\prime})} & = &
\sum_{\begin{array}{c}{\rm roots} \, \mu_k \\ {\rm of}\,  p(N)\end{array}}
\frac{\log(\mathcal{N}-\mu_k) -\log(N_i-\mu_k)}{3 d \mu_k^2 + 2 c \mu_k + b} \, .
\end{eqnarray}

With the parameter values of the problem, eqs.~(\ref{coeffa}--\ref{coeffd}),
only one root (which we will denote $\mu_1$) of $p(N)$ is real, while the
other two are complex conjugate of each other
\begin{eqnarray}
\mu_3 & = & \mu_2^{\star} \, , \\
& = & \mathfrak{a} - i \mathfrak{b} \, .
\end{eqnarray}
Defining 
\begin{eqnarray}
\label{defz}
z & = & \mathcal{N} - \mu_2 \, , \\
\label{defzstar}
z^{\star} & = & \mathcal{N} - \mu_3 \, ,
\end{eqnarray}
and equivalently for $z_i$, we expand the denominator of the formal solution
(r.h.s. of eq.~(\ref{formalsol}))
\begin{eqnarray}
\nonumber
3 d \mu_k^2 && + 2 c \mu_k + b = \\ 
\label{def-poly}
&&[3d(\mathfrak{a}^2-\mathfrak{b}^2)  + 2 c
\mathfrak{a} + b] \pm i [6 d \mathfrak{a} \mathfrak{b} + 2 c \mathfrak{b}] \, ,
\\
\label{def-Omegas}
& \equiv & \Omega_r \pm i \Omega_i \, , \\
k & = & 2,3
\end{eqnarray}
where the $+$ ($-$) sign in eqs.~(\ref{def-poly},\ref{def-Omegas}) corresponds
to $k=2$ ($k=3$).

Thus, the contributions to eq.~(\ref{formalsol}) coming from the two complex
roots can be rewritten as
\begin{eqnarray}
\nonumber
\sum_{\mu_k, k=2,3} && \frac{\log(\mathcal{N}-\mu_k) -\log(N_i-\mu_k)}{3 d
\mu_k^2 + 2 c \mu_k + b} = \\
\nonumber
&& \left( \log(z) - \log(z_i) \right) \frac{\Omega_r - i \Omega_i}{|\Omega|^2} +
\\
&&\left( \log(z^{\star}) - \log(z_i^{\star}) \right) \frac{\Omega_r + i
\Omega_i}{|\Omega|^2} \, ,
\end{eqnarray}
with $z_i \equiv N_i - \mu_2$ (cf. eqs.~(\ref{defz},\ref{defzstar})).

Substituting into the full expression, we finally obtain the solution for
eq.~(\ref{genintegral}):
\begin{eqnarray}
\nonumber
t & = & \frac{\log(\mathcal{N}-\mu_1) -\log(N_i-\mu_1)}{3 d \mu_1^2 + 2 c \mu_1
+ b} + \\
\nonumber
&& 2 \left[\log|z| - \log|z_i| \right] \frac{\Omega_r}{|\Omega|^2} + \\
\label{Nsolution}
&& 2\left[ \operatorname{Arg} \left\{ z \right\} - \operatorname{Arg} \left\{
z_i \right\} \right] \frac{\Omega_i}{|\Omega|^2} \, .
\end{eqnarray}

This relationship implicitly defines the approximate solution for the carrier
number and gives the most practical way of numerically representing the
analytical solution:  treating time as the dependent variable,
$t(\mathcal{N})$, the function can be straightforwardly plotted (exchanging
the horizontal and vertical axes allows one to restore the $\mathcal{N}(t)$
appearance of the function).  Alternately, simple handling allows for a
mathematical expression for $\mathcal{N}(t)$ which takes the form:
\begin{eqnarray}
\nonumber
&&\left( \mathcal{N}(t) - \mu_1 \right) \left( \mathcal{N}^2(t) - 2 \Re e
\left\{ \mu_2 \right\} \mathcal{N}(t) + |\mu_2|^2 \right) \\
&& \qquad \qquad \nonumber e^{2 \operatorname{Arg} \left\{ \mathcal{N}(t) -
\mu_2 \right\} } = \\ 
\label{expl-an-sol}
&& \qquad \qquad \mathcal{K}
 e^t \, , 
\end{eqnarray}
where
\begin{eqnarray}
\nonumber
\mathcal{K} & = & \left( N_i - \mu_1 \right) \\
\label{expl-an-coeff}
& & \left( N_i^2 - 2 \Re e \left\{ \mu_2 \right\} N_i +
|\mu_2|^2 \right) e^{2 \operatorname{Arg} \left\{ N_i - \mu_2 \right\}
} \, ,
\end{eqnarray}
which can be used as an alternative definition of
$\mathcal{N}(t)$~\cite{alt-def-N}. 

\begin{figure}[ht!]
\includegraphics[width=0.9\linewidth,clip=true]{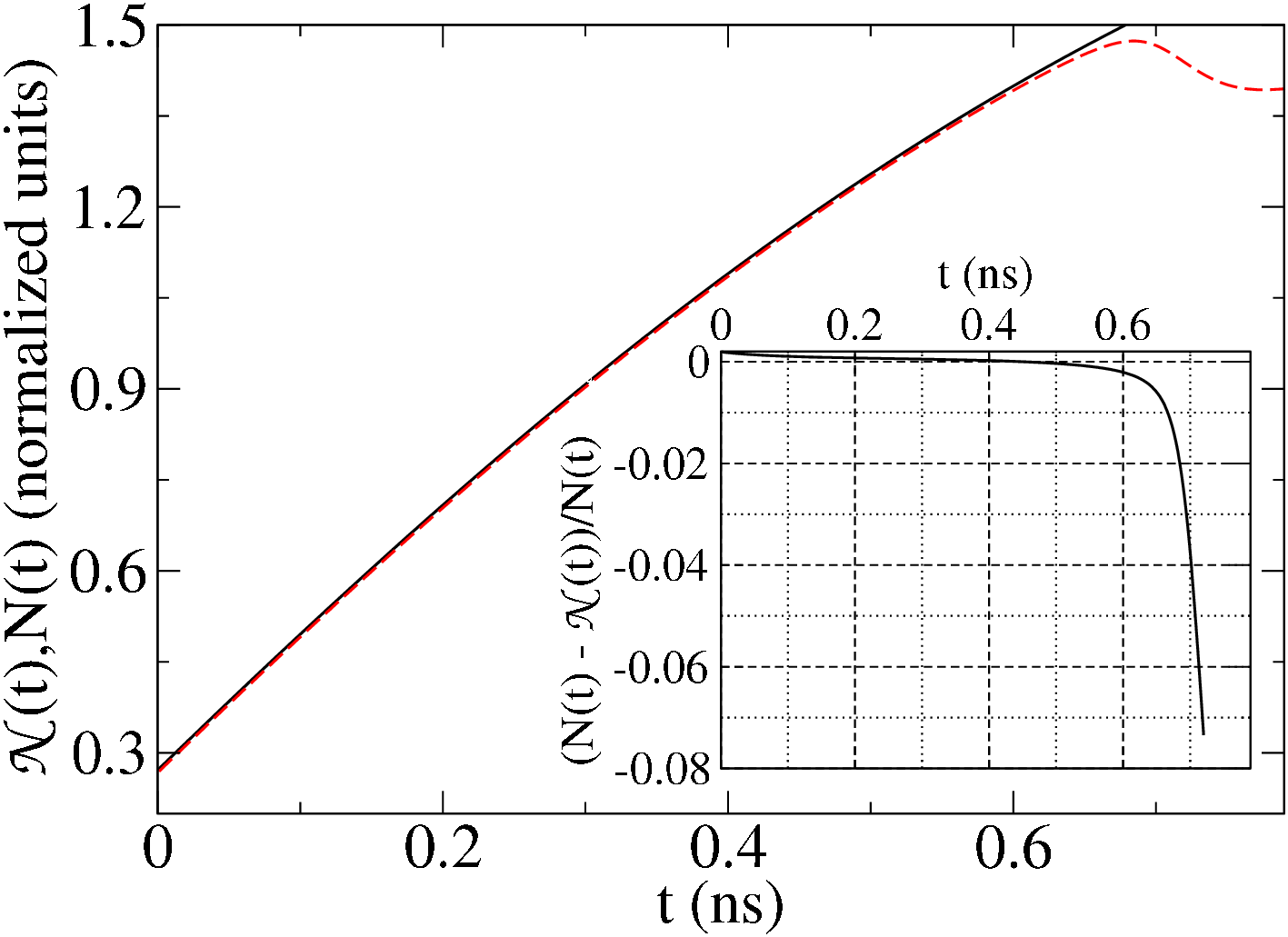}
\caption{
Comparison between the approximate analytical solution (solid line)
$\mathcal{N}(t)$ and the complete carrier number (dashed line) $N(t)$
resulting from the integration of the full model, eqs.~(\ref{model}).  The
inset shows the difference between the full solution $N(t)$ and the analytical
solution $\mathcal{N}(t)$ until the latter diverges away (at $t \approx 0.72
ns$).  Notice that even extremely near the intensity peak (cf.
Fig.~\ref{time-trace}) the difference between the approximate and full
solution remains modest ($\lesssim 8\%$).
} 
\label{sols-N} 
\end{figure}

Fig.~\ref{sols-N} compares the analytical solution $\mathcal{N}(t)$ to the
computed $N(t)$ obtained from the numerical integration of the full system,
eqs.~(\ref{model}).
The agreement between the approximate, analytical solution and the full
integration is excellent up until values of time which are quite close to the
intensity peak (occurring at $t \approx 0.72 s$), as shown in the inset.  The
deviation is initially positive and remains below 0.5\% (in absolute value)
until $t \approx 0.62 ns$.  The initial positivity of the relative
deviation,$\frac{N(t)-\mathcal{N}(t)}{N(t)}$, resides in the fact that
the
full solution requires a somewhat larger initial carrier number to support the
different field modes, while in the approximate form no energy is lost to
support them.  The switch in sign in the difference between the exact and the
approximate value for the carrier number comes from the fact that once the
different lasing modes grow sufficiently large, the carrier number saturates,
instead of continuing its growth, as in the approximate expression.  The
difference curve (inset) is numerically computed  by integrating the full set
of equations, eqs.~(\ref{model}), and a set equivalent to the analytical
solution, obtained by removing from eq.~(\ref{modelb}) the last term $\left(
\sum_j \Gamma G_j(N) I_j \right)$~\cite{check-an-num}.

Notice that at $t \equiv t_{th} \approx 0.62 ns$, the carrier number $N(t)$
crosses its asymptotic value ($N(t_{th}) = \overline{N} \equiv N(t \rightarrow
\infty)$, cf. Fig.~\ref{time-trace}).  In a laser model without spontaneous
emission such a value would define the laser threshold, separating the
decoupled dynamics for the population inversion (carrier number) from the
strongly coupled regime (above threshold)~\cite{Narducci}.  Here, this
value sets the upper limit to the carrier number for which we can consider its
dynamics decoupled from that of the modal intensities (when growing out of
noise), thus we can extend the concept of ``threshold" even in the
presence of spontaneous emission.

\subsection{Decoupled modal dynamics}\label{decoupled-modes}

We begin this section with an overview of the modal dynamics during the
transient evolution, regardless of the degree of coupling with the carrier
number.  This allows us to get a general picture, which we then refine in the
analysis of the decoupled regime (this section) and of the strongly coupled
one (next section).

The individual mode traces in Fig.~\ref{time-trace} show that at short times
the lateral modes contribute more than their steady state share.  Their peak
(cf., e.g., curves (f) and (g) corresponding to modes 5 and 7 places away from
line center) is much higher than the asymptotic intensity towards which they
tend, contrary to those modes close to line center (for the central mode even
the peak intensity is lower than its steady state value)~\cite{slowmodes}.
This strongly suggests that a mechanism must exist for these modes to grow, in
transient, well beyond their final value, while the very few central modes do
not.  Indeed, the accumulated intensity of the modes farthest away in the
wings can exceed, in transient, that of the central mode, thus highlighting
the impact that the side modes have on the initial phases of the dynamics.

\begin{figure}[ht!]
\includegraphics[width=0.9\linewidth,clip=true]{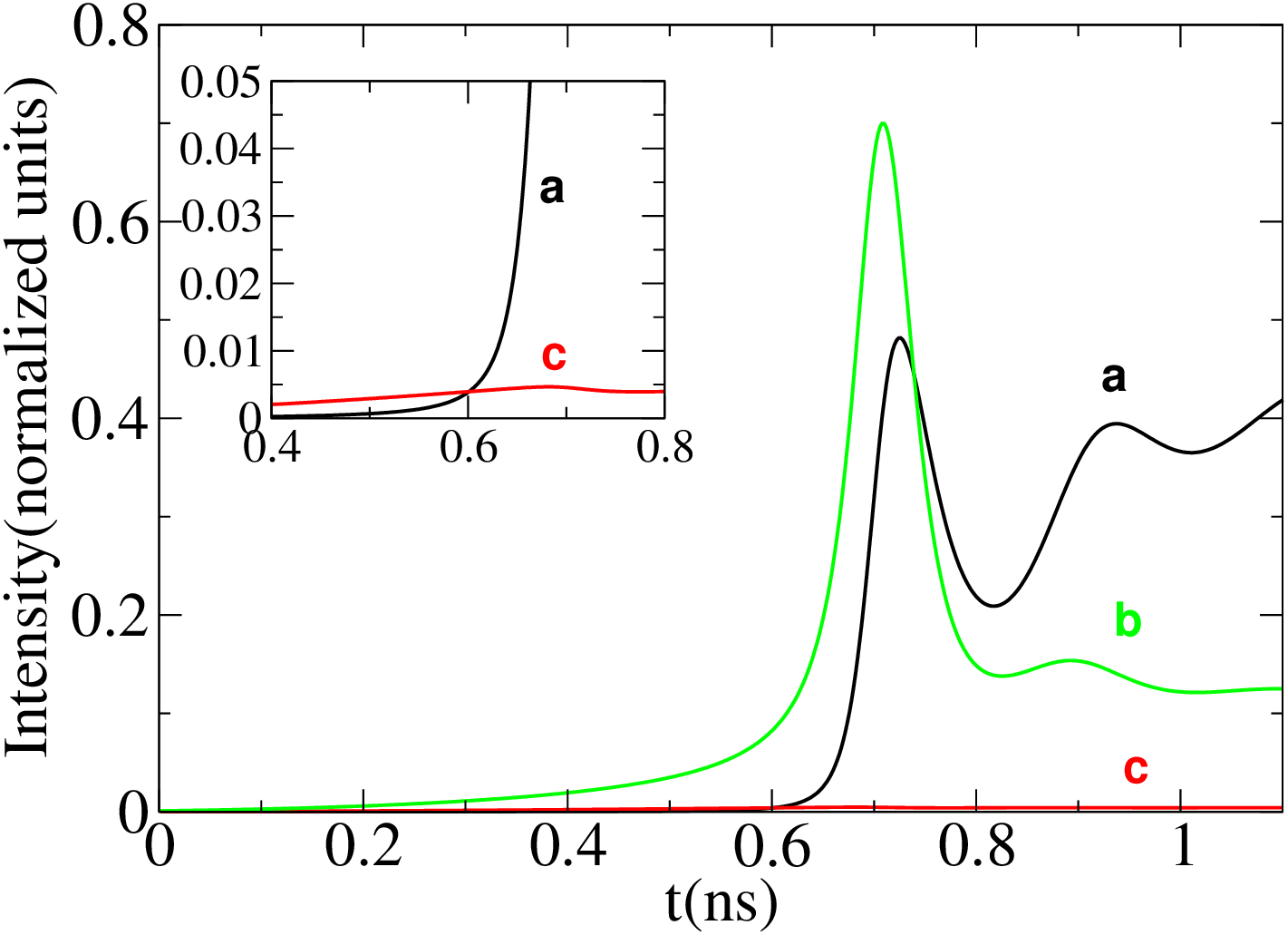}
\caption{
Temporal evolution of modes or groups of modes.  Curve (a) traces the
evolution of the central mode, (c) the sum of the most distant 20 modes on
each side of line center, $I_{far}$, (cf. text for details) and (b) the sum of
all side modes, $I_{M-17}$, save for those near line center (cf. text
for details).  The inset shows the initial phases of the transient and shows
that the sum of the far modes (c) exceeds the intensity of the central mode
(a) up until $t \approx t_{th}$.
}
\label{t113-sum} 
\end{figure}

A more complete illustration of this point is given by Fig.~\ref{t113-sum},
which compares the time-dependent intensity of the central mode -- curve (a)
--, to the cumulative contribution of the most distant twenty modes (on each
side, i.e. $I_{far} = \sum_{j=1}^{20} I_j + \sum_{j=94}^{113} I_j$) -- curve
(c)), and to that of all modes save for the 17 central ones (i.e.,$I_{M-17}
= \sum_{j=1}^{48} I_j + \sum_{j=66}^{113} I_j$ -- curve (b)). 

The inset shows that below threshold $I_{far}$ (c) gives a contribution larger
than that of the central mode (a) -- until $t \lesssim 0.6 ns$ -- and that
after a small maximum, at $t \approx 0.68 ns$, it drops down again.  Thus, the
influence of the very far modes is of some importance only in the very first
phases of the dynamics.  Considering, though, that these modes are very far
out in the wings and that their asymptotic value is order of magnitudes lower
than that of the central mode, their strong influence on the dynamics up to
threshold is an indicator of the essential difference between the transient
intensity distribution and the asymptotic one!  

The picture changes more dramatically if we include the lateral modes,
$I_{M-17}$, which carry little energy at steady state (globally of the order
of 3\% of the total intensity), but which in transient give a cumulative peak
larger than the one of the central mode (curve (b)).  This contribution is
rather striking.  Indeed, it displays a faster growth than the central mode
and deformed (incomplete) oscillations. The latter are explained by the fact
that the effective relaxation frequency is not constant over the whole set of
modes (cf.~\cite{Marcuse1983} for discussion on a slightly different model and
Section~\ref{strong-coupling}, in this paper), thus the sum over the ensemble
deforms and attenuates the oscillations.

The substantial contribution of $I_{M-17}$ to the peak in transient is a
consequence of the observation that in the initial phases of the dynamics the
lateral modes contribute, relatively, much more than they do at steady state
(or at least on the long time scales).  In addition, we remark that the
anticipated peak for $I_{M-17}$ contributes to a faster growth of the total
intensity.  We now analyze the initial portion of the transient, i.e., the
decoupled regime.

The remarks concerning the transient leading to, but excluding, the intensity
peaks (and oscillations) can be better understood with the help of the
following approximate analysis.  Using the analytical (approximate) solution
for $\mathcal{N}(t)$ (eq.~(\ref{Nsolution})) we can now decouple the modal
intensities to obtain an approximate dynamical evolution:
\begin{eqnarray}
\nonumber
\frac{d \, {^dI_j}}{d t} & \approx & \left[ \Gamma G_j(\mathcal{N}) -
\frac{1}{\tau_p}
\right] {^dI_j} + \\
\label{decoupledI}
&& \beta_j B \mathcal{N}(t) \left[ \mathcal{N}(t) + P_0 \right] \, ,
\end{eqnarray}
where the superscript $^d$ denotes the approximate, decoupled variable.

The following auxiliary quantities simplify the notations:
\begin{subequations}\label{defcoeffjs}
\begin{eqnarray}
\alpha_j & = & \Gamma g_p \left[ 1 - 2 \left( \frac{\lambda_j -
\lambda_p}{\Delta \lambda_G} \right)^2 \right] \, , \\
\mathcal{K}_j & = & \alpha_j N_0 + \frac{1}{\tau_p} \, ,\\
B_j & = & \beta_j B \, ,
\end{eqnarray}
\end{subequations}
where we have neglected the saturation term $\epsilon I_t$ in the expression
for $G_j(N)$, eq.~(\ref{coeff2}), since we are examining the transient
portions in which the intensity is very small, $\epsilon$ being also very
small (cf.  Table III in~\cite{slowmodes}).  This way, the modal intensities
obey the set of M decoupled ODEs
\begin{eqnarray}
\label{simpleI}
\frac{d \, {^dI_j}}{dt} & \approx & \left[ \alpha_j \mathcal{N}(t) -
\mathcal{K}_j
\right] {^dI_j} +B_j \mathcal{N}(t) \left[ \mathcal{N}(t) + P_0 \right] \, ,
\end{eqnarray}
which can be globally analyzed.

Over very short time intervals $\Delta t$ ($\Delta t$: $\mathcal{N}(t + \Delta
t) \approx \mathcal{N}(t)$, $\forall \, t:$ ($0 \le t \lesssim t_{th}$) )
we can consider the two terms $\left[ \alpha_j \mathcal{N}(t) - \mathcal{K}_j
\right]$ and $B_j \mathcal{N}(t) \left[ \mathcal{N}(t) + P_0 \right]$ as being
constants, thus reducing eq.~(\ref{simpleI}) to an ensemble of ODEs with
constant coefficients.  In such a case, formal integration provides
immediately a solution for the approximate problem, eq.~(\ref{simpleI}):
\begin{eqnarray}
\nonumber
^dI_j(t_k + t^{\prime}) & \approx & -\frac{B_j \mathcal{N}(t_k) \left[
\mathcal{N}(t_k) + P_0 \right]
}{\alpha_j \mathcal{N}(t_k) - \mathcal{K}_j} + \\
\nonumber
& & \left[ {^dI_j(t_k)} + \frac{B_j \mathcal{N}(t_k) \left[
\mathcal{N}(t_k) + P_0 \right]
}{\alpha_j \mathcal{N}(t_k) - \mathcal{K}_j} \right] \\
\label{anal-int-shtime}
& & e^{\left[ \alpha_j \mathcal{N}(t_k) - \mathcal{K}_j \right]
t^{\prime}} \, , \\
\nonumber
& = & -\frac{\mathcal{B}_j(t_k)}{\gamma_j(t_k)} +  \\
\label{simpler-approx-form}
& & \left[ {^dI_j(t_k)} + \frac{\mathcal{B}_j(t_k)}{\gamma_j(t_k)} \right]
e^{\gamma_j(t_k) \cdot t^{\prime}} \, ,\\
\label{def-gamma-j}
\gamma_j(t_k) & \equiv &  \alpha_j \mathcal{N}(t_k) - \mathcal{K}_j \quad (<
0) \, , \\
\label{def-beta-spont-em}
\mathcal{B}_j(t_k) & \equiv & B_j \mathcal{N}(t_k) \left[
\mathcal{N}(t_k) + P_0 \right]
\, , \\
t_k & \equiv & k \times \Delta t \, , \\
k & = & 0 \ldots q \, , \\
0 \le & t^{\prime} & \le \Delta t \, , \\
{^dI_j(t_0)} & = & I_j(t_0) \, ,
\end{eqnarray}
are the initial conditions for each modal intensity.
The inequality in parenthesis (eq.~(\ref{def-gamma-j})) holds strictly in
the range of validity of the current approximation.  Here $t^{\prime}$ is a
{\it local} time defined in each interval $\Delta t$ and is zeroed at the
beginning of each time step.

\begin{figure}[ht!]
\includegraphics[width=0.9\linewidth,clip=true]{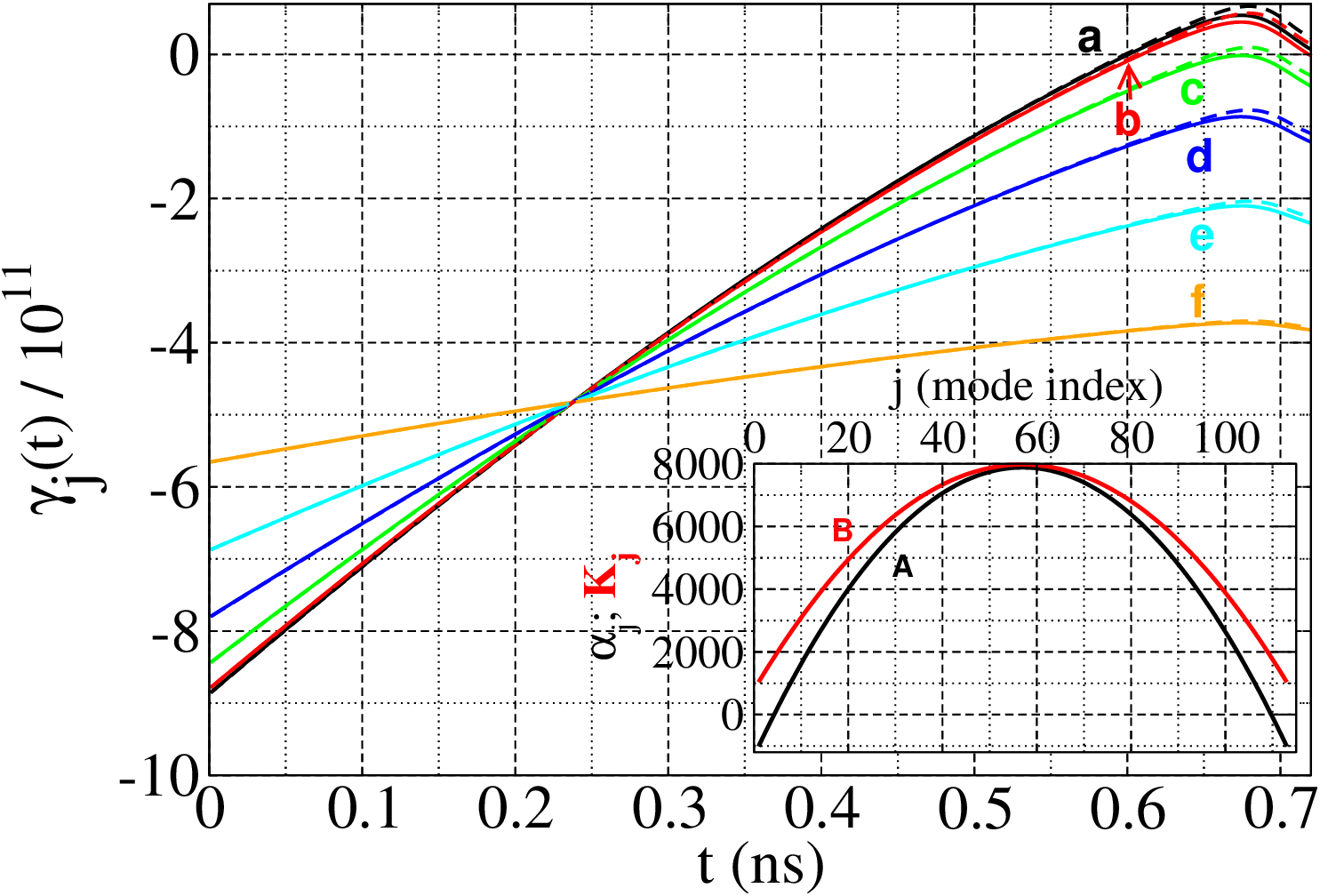}
\caption{
Time evolution of $\gamma_j$ (in units of $10^{11}$) for selected modes: (a) j
= 57; (b) j = 50; (c) j = 40; (d) j = 30; (e) j = 20; (f) j = 10.  The solid
lines correspond to the values of $\gamma_j$ computed from the full set of
equations, eqs.~(\ref{model}), while the dashed lines (which separate only for
$t > t_{th}$ from the solid lines) are computed for the analytical solution
$\mathcal{N}(t)$, eq.~(\ref{expl-an-sol},\ref{expl-an-coeff}).  The inset
shows the modal dependence of $\alpha_j$ (curve (A)) and $\mathcal{K}_j$
(curve (B)).  For graphical purposes we have plotted instead of
$\mathcal{K}_j$ its normalized and shifted version {\bf K}$_j =
\frac{\mathcal{K}_j}{1 \times 10^8} - 3 \times 10^3$.  
} 
\label{gamma-j}
\end{figure}

The coefficients $\mathcal{B}_j(t_k)$ and
$\gamma_j(t_k)$~\cite{diff-gamma-beta} are constant over a small time
interval $\Delta t$ and are ``updated", as time evolves, at the next
value $t_k$.  This, in
itself, is not a limitation since their functional dependence is known and
explicitly given by
eqs.~(\ref{expl-an-sol},\ref{expl-an-coeff},\ref{defcoeffjs}).  Thus,
eq.~(\ref{anal-int-shtime}) is a ``piecewise", recursive solution over a
discrete
ensemble of times $t_k$ ($k = 0 \ldots q \, : t_0 = 0, t_q \approx t_{th}$).
At each time step, each $^dI_j$ starts to converge, for a short time, towards
its locally asymptotic solution before being updated to the next time step.
The form of the approximate solution, eq.~(\ref{anal-int-shtime}), highlights
the role of the different coefficients:  the spontaneous emission term,
$\mathcal{B}_j$, controls the amplitude (together with $\gamma_j$ in the
prefactor of eq.~(\ref{simpler-approx-form})), while gain, $\alpha_j$, and
cavity losses, $\mathcal{K}_j$, combined into $\gamma_j$ play the role of an
{\it effective relaxation constant}.  Notice that
$\gamma_j$ results from the composition, eq.~(\ref{def-gamma-j}), of the two
functions shown in the inset of Fig.~\ref{gamma-j} (notice the shift and
multiplicative factor for curve B -- cf. figure caption) which represent
$\alpha_j$
(curve A) and $\mathcal{K}_j$ (curve B).  The different local curvatures,
combined with the multiplication of curve A by the carrier number $N(t)$ (or
equivalently 
$\mathcal{N}(t)$ when using the approximate solution), differentiate the
coefficient $\gamma_j$ for the individual modes (with the possibility of
obtaining, in transient, positive values when curve A$\times N$ is larger than
curve B -- possible only outside the range of validity of the current
approximation).

\begin{figure}[ht!]
\includegraphics[width=0.9\linewidth,clip=true]{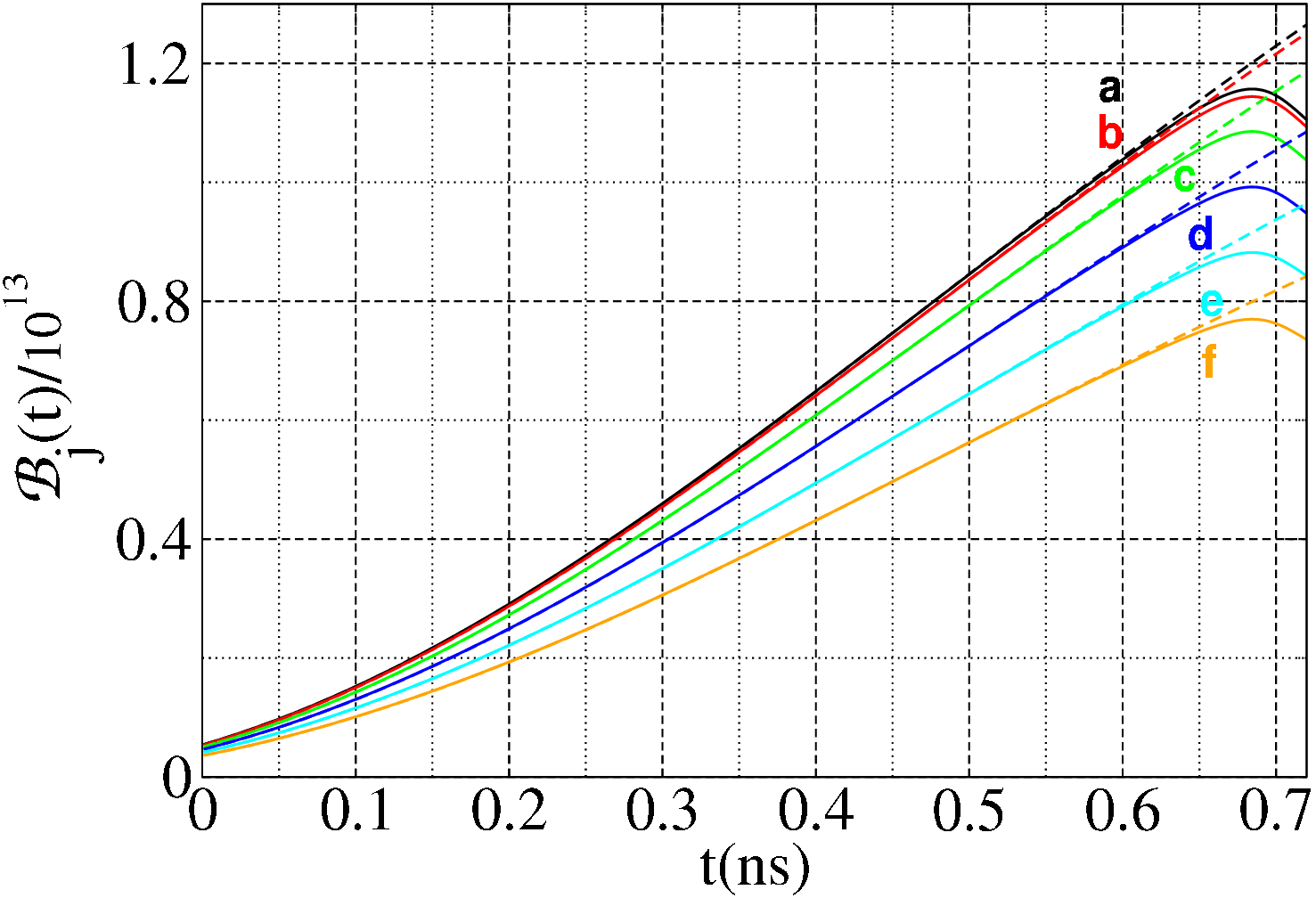}
\caption{
Time evolution of the spontaneous emission contribution to each individual
mode (in units of $10^{13}$) for selected modes: (a) j = 57; (b) j = 50; (c) j
= 40; (d) j = 30; (e) j = 20; (f) j = 10.  The solid lines correspond to the
values of $\beta_j N(t) \left[ N(t) + P_0 \right]$
computed from the full set of equations, eqs.~(\ref{model}), while the dashed
lines (which separate only for $t > t_{th}$ from the solid lines) are computed
for the analytical solution $\mathcal{N}(t)$ in place of $N(t)$,
eq.~(\ref{expl-an-sol},\ref{expl-an-coeff}).  The labels (a) and (b)
correspond, from top to bottom, to the top two sets of curves (solid and dashed,
respectively) which can be distinguished only on the far right of the figure.
} 
\label{beta-j} 
\end{figure}

In order to better understand the behaviour of this approximate solution, we
trace the coefficients $\gamma_j$ and $\mathcal{B}_j$ for selected modes in
Figs.~\ref{gamma-j} and \ref{beta-j}, respectively.  Fig.~\ref{gamma-j} shows
the evolution of $\gamma_j$ as a function of time for a
representative sample of modes.  Aside from the initial portions of the
transient~\cite{transp} ($t \lesssim 0.24 ns$) the larger $\gamma_j$, the
closer
is the corresponding mode to line center.  Since $\gamma_j$'s behave as
effective relaxation constants, the figure immediately shows that the wing
modes, with their larger (negative) values of $\gamma$, react on a shorter
time scale than those modes close to line center.  This explains the numerical
observation of Fig.~\ref{t113-sum}, where the wing modes grow faster than the
central one, and partly accounts for the excess growth of these modes beyond
their asymptotic values~\cite{constr-bth}.  The constraint which forces the
energy to be distributed over all modes (cf. discussion in Section IVA
of~\cite{slowmodes}), transfers the excess which cannot be taken up by the
stronger (and slower) modes to the faster wing modes during the transient.  On
the longer time scales, on which the central mode(s) react(s), the energy is
transferred back, allowing the side modes to relax to their asymptotic values.
Notice that the agreement between the coefficients calculated from the
analytical solution for the carrier number and from the full numerical
solution holds extremely well until after threshold:  in Figs.~\ref{gamma-j}
and~\ref{beta-j} one can distinguish solid and dashed curves only in the top
right hand portion of the figure. 

In addition to the component ($\gamma_j \, {^dI_j}$ in eq.~(\ref{simpleI}), with
the substitutions, eqs.~(\ref{def-gamma-j},\ref{def-beta-spont-em}))
discussed so far, the
temporal
evolution of the individual modes includes the contribution of the spontaneous
emission $\mathcal{B}_j$ whose amplitude
is plotted in Fig.~\ref{beta-j} (dashed lines) for a subset of modes, as a
function of time
(thus, of carrier number, cf.~eq.~(\ref{Nsolution})) and is compared to the
amplitude (solid lines) for the full model, eqs.~(\ref{model}).  Its main
feature is the
similarity in its contribution to all modes (maximum deviation, less than a
factor 2) even at the end of the time interval (largest $\mathcal{N}(t)$).
The bundle of curves opens up with time, increasing by about one order of
magnitude for the central mode ($j = 57$) -- somewhat less for the plotted
mode farthest away from line center ($j=10$).  For the latter mode we remark
that $\frac{\gamma_{10}(t_{th})}{\gamma_{10}(0)} \approx \frac{2}{3}$, thus
maintaining a nearly constant, fast relaxation throughout the decoupled
regime.  The fast, nearly constant, reaction time explains the sizeable growth
of the wing modes, in spite of the smaller contribution which they receive
from the spontaneous emission.  The moderate change in $\gamma_{10}$ (and
equivalent from the wing modes) ensures a slight growth in the prefactor of
eq.~(\ref{simpler-approx-form}), thus a rather stable energy contribution of
these modes below threshold.  For comparison, in the units of
Fig.~\ref{gamma-j}, $\gamma_{57}(t \rightarrow \infty) \approx - 10^{-3}$
(cf.~\cite{slowmodes}).  Hence, the effective relaxation rate changes
considerably for the center mode and is eventually responsible for their
predominance at the end of the below-threshold regime, even though the side
modes maintain a level which is considerably larger than their asymptotic
contribution.

\begin{figure}[ht!]
\includegraphics[width=0.9\linewidth,clip=true]{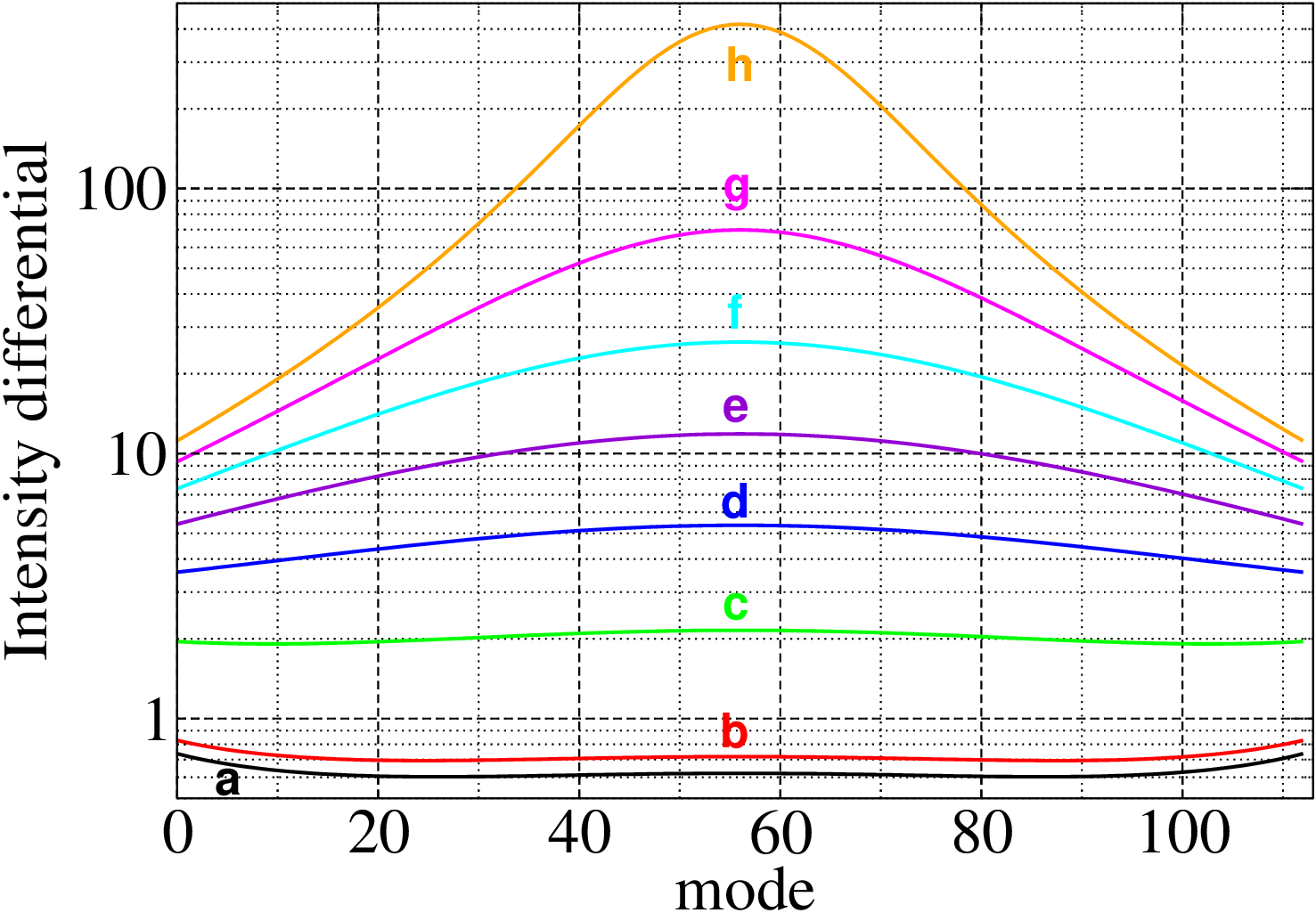}
\caption{
Differential increment (numerically computed with a time step $\Delta t = 1
\times 10^{-14} s$) for the individual mode intensities (r.h.s. of
eq.~(\ref{simpleI})) plotted at different times:  (a) $t = 0.001ns$
($\mathcal{N} = 2.17558 \times 10^7$); (b) $t = 0.01ns$ ($\mathcal{N} =
2.893353 \times 10^7$); (c) $t = 0.1ns$ ($\mathcal{N} = 4.903313 \times
10^7$); (d) $t = 0.2ns$ ($\mathcal{N} = 7.033165 \times 10^7$); (e) $t =
0.3ns$ ($\mathcal{N} = 9.024550 \times 10^7$); (f) $t = 0.4ns$ ($\mathcal{N} =
1.085107 \times 10^8$); (g) $t = 0.5ns$ ($\mathcal{N} = 1.249524 \times
10^8$); (h) $t = 0.6ns$ ($\mathcal{N} = 1.394853 \times 10^8$).  
}
\label{I-incr} 
\end{figure}

The approximate solution, eq.~(\ref{simpler-approx-form}), shows the role played
by $\frac{\mathcal{B}_j(t_k)}{\gamma_j(t_k)}$, which is nothing else than
the ratio between the coefficients shown in Figs.~\ref{gamma-j},\ref{beta-j}.
Thus, another way of interpreting the dynamics below threshold is to plot,
Fig.~\ref{I-incr}, the r.h.s. of eq.~(\ref{simpleI}) -- thus the intensity
increment -- for the whole ensemble of modes at selected times (i.e., for
selected values of $\mathcal{N}(t)$), from the initial condition up to
threshold.

When the carrier number is well below transparency (curves a-c) the gain is
nearly equal for all modes, with a somewhat higher level in the wings for the
first two curves~\cite{winggain}.  This implies that the modes far from line
center start growing faster than the central ones.  Near transparency, curve
(d), the central modes start dominating and the combined effect of their
somewhat larger intensity and of their increasing coefficients (cf.
Figs.~\ref{gamma-j},\ref{beta-j}) favors them.  For the parameters of curve
(e) the laser is already beyond transparency ($N_0 = 7.8 \times 10^7$, $t
\approx 0.24 ns$) and the tendency for a strong differential growth becomes
obvious and sharpens itself with growing values of $\mathcal{N}$.  Equivalent
curves showing the ratio (cf. eq.~(\ref{simpler-approx-form}))
$\frac{\mathcal{B}_j}{\gamma_j}$ (each component plotted in
Figs.~\ref{gamma-j},\ref{beta-j}) look qualitatively very similar to those of
Fig.~\ref{I-incr}, since they differ only for the intensity $^dI_j$ which
multiplies $\gamma_j$ (the relative deformation is most visible for curve (h),
which is more strongly reshaped by the larger modal intensity variations).

\begin{figure}[ht!]
\includegraphics[width=0.9\linewidth,clip=true]{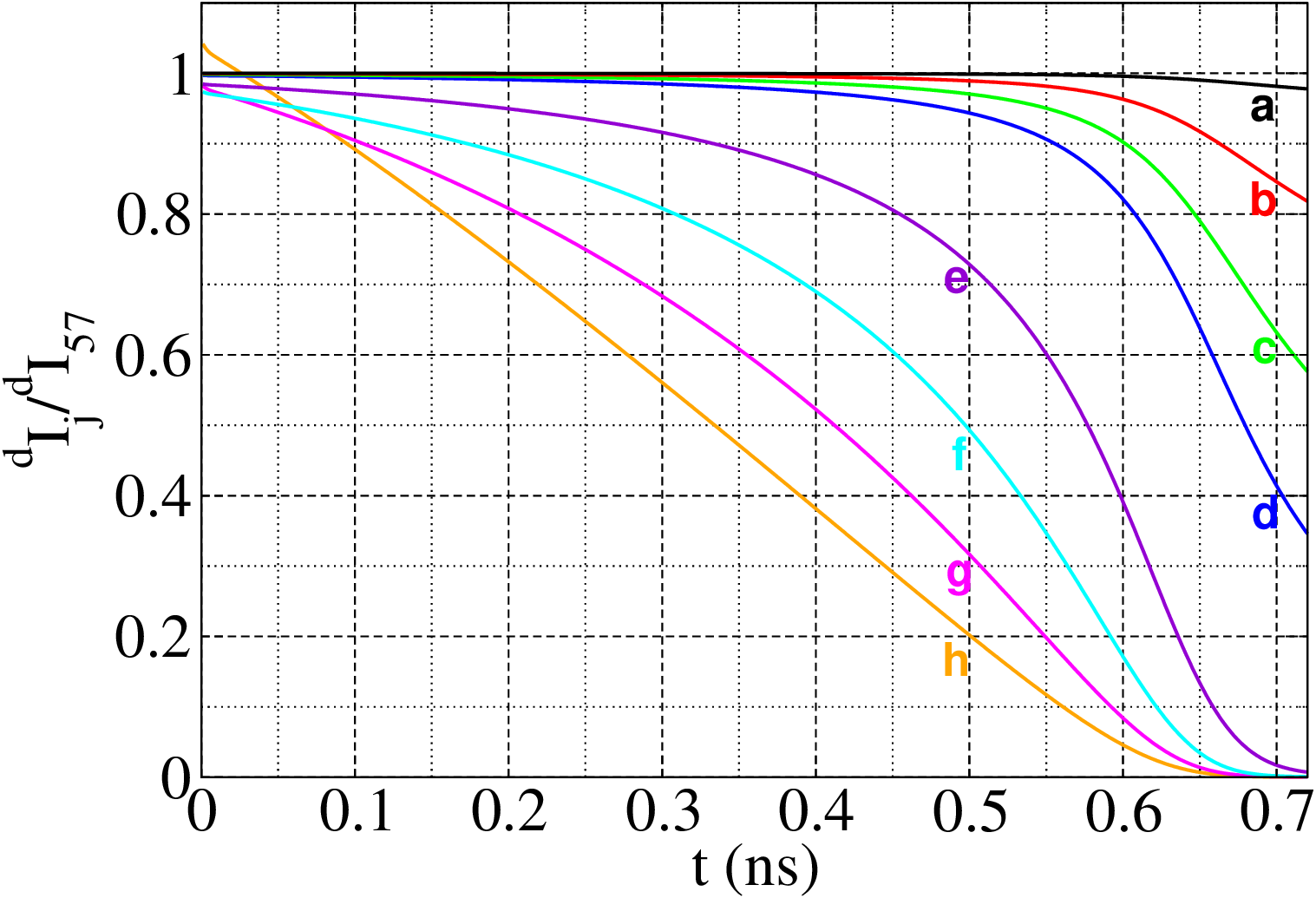}
\caption{
Intensity ratio $\frac{^dI_j}{^dI_{57}}$ for a selected subset of modes:  (a) $j
= 56$; (b) $j=54$; (c) $j=52$; (d) $j=50$; (e) $j=40$; (f)
$j=30$; (g) $j=20$; (h) $j=10$.  The ratios are computed with the decoupled
system of equations, eqs.~(\ref{Nsolution},\ref{simpleI}).
} 
\label{ratio-int} 
\end{figure}

In spite of the clear increase in differential intensity growth between the
central and the side modes displayed in Fig.~\ref{I-incr} the ratio remains
finite up until threshold, as shown in Fig.~\ref{ratio-int}, where the
amplitude of each intensity mode is displayed normalized to the intensity of
the central mode.  Fig.~\ref{ratio-int} shows again that the gain is largest
for the wing modes at the start of the transient (curve (h), $j=10$, is the
highest at $t = 0$), due to
their lower absorption and that during the whole evolution in the decoupled
regime the relative weight of the side modes largely exceeds their asymptotic
value $\left( \left. \frac{^dI_{10}}{^dI_{57}} \right|_{t \approx 0.6 ns}
\approx
0.05 \right.$ -- cf. Fig.~\ref{ratio-int} -- to be compared to
$\frac{\overline{I}_{10}}{\overline{I}_{57}} \approx 1.5  \times 10^{-4}$ from
Table~\ref{max-time}{\Large )}.  This result is consistent with the numerical
observation -- obtained from the integration of the full model,
eqs.~(\ref{model}) -- showing that at $t = 0.6 ns$ the cumulative intensity of
the 20 modes farthest away from line center, $I_{far}$, exceeds that of the
central mode (cf. Fig.~\ref{t113-sum}). It also explains why at $t =
\overline{t}$ the intensity distribution is far from the asymptotic one
(Fig.~\ref{time-trace} and~\cite{slowmodes}):  the wing modes at the beginning
of the strongly coupled regime hold a very large relative amount of energy
($\approx$ 300
times its relative asymptotic value for $j = 10$).

Besides the numerical evidence of Fig.~\ref{ratio-int},
eq.~(\ref{simpler-approx-form}) shows the reason for the not-so-large
differences among modes, below threshold.  Expanding the exponential in
eq.~(\ref{simpler-approx-form}) to first order, we obtain an approximate form
for the modal intensity:
\begin{eqnarray}
^dI_j(t_k+t^{\prime}) & \approx & {^dI_j(t_k)} \left[ 1 + \gamma_j(t_k) \cdot
t^{\prime} \right] + \mathcal{B}_j(t_k) \cdot t^{\prime} \, ,
\end{eqnarray}
thus its increment relative to the previous time step takes the form:
\begin{eqnarray}
\nonumber
^dI_j(t_k+t^{\prime}) - {^dI_j(t_k)} & \approx & {^dI_j(t_k)} \gamma_j(t_k)
\cdot t^{\prime} + \\
\label{incrementIj}
& & \mathcal{B}_j(t_k) \cdot t^{\prime} \, .
\end{eqnarray}
The first term on the r.h.s. of eq.~(\ref{incrementIj}) represents a relaxation
($\gamma_j < 0$, $\forall (j,t_k)$), while the second one is the modal
intensity increment brought about by the (average) spontaneous emission
contribution.  Thus, the differential growth between center and wing modes
remains moderate (cf. Fig.~\ref{beta-j}), which explains why even close to
threshold the ratio of the wing modes to the center mode (Fig.~\ref{ratio-int})
remains very large compared to its asymptotic value.  Notice that
eq.~(\ref{incrementIj}) indirectly determines a time step $\Delta t$, since
$\Delta t \ll \frac{1}{\mathcal{B}_j(t_k)}$, $\forall (j,
t_k)$~\cite{convergence-radius}.

\begin{figure}[ht!] 
\includegraphics[width=0.9\linewidth,clip=true]{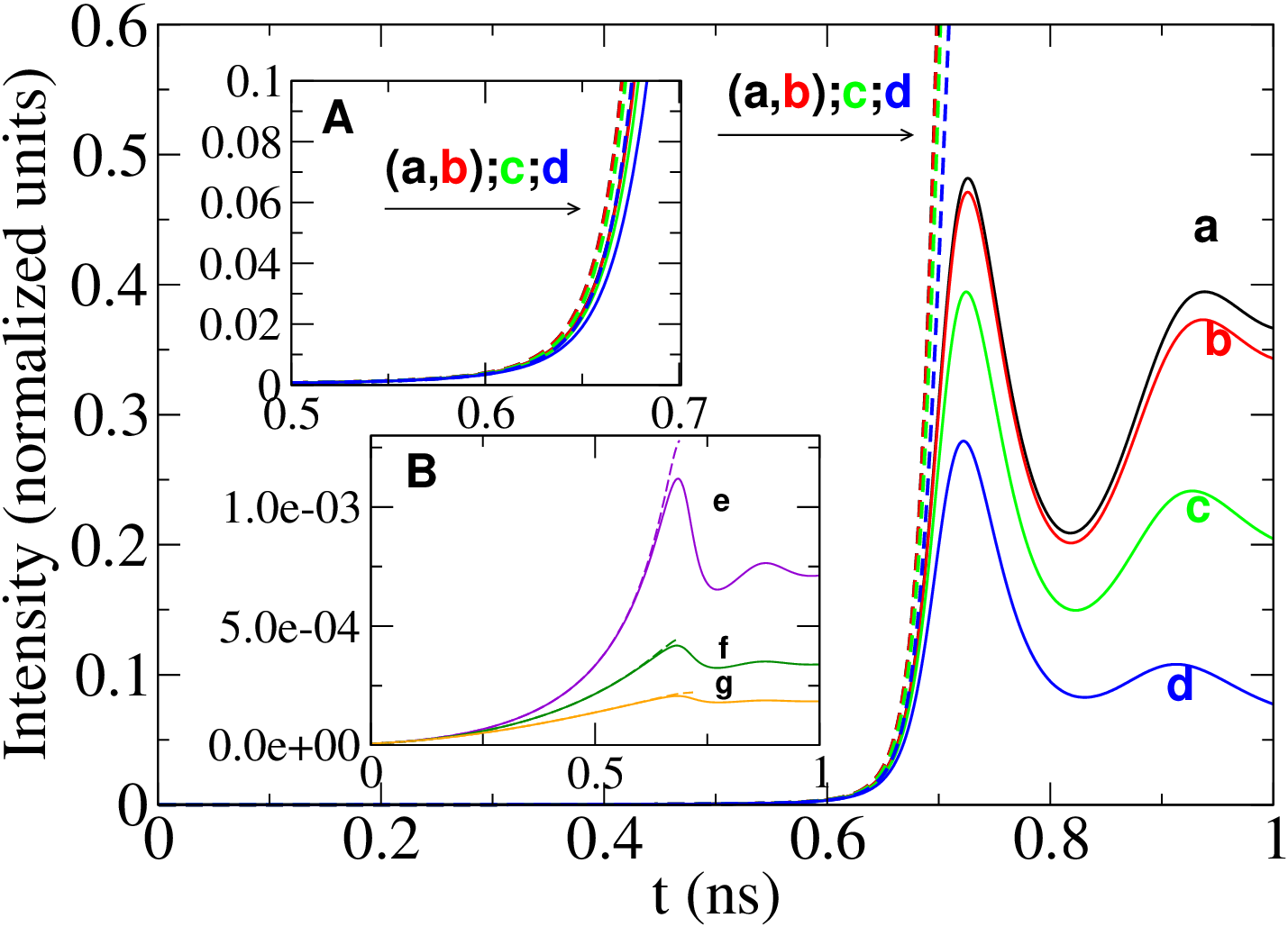}
\caption{
Temporal evolution of the intensity for a selected subset of modes:  (a)
$j=57$; (b) $j = 56$; (c) $j=54$; (d) $j=52$; (e) $j=30$; (f) $j=20$; (g)
$j=10$.  The solid lines display the result of the integration of the full
model, eqs.~(\ref{model}), while the dashed lines show, for comparison, the
corresponding intensity values calculated from the decoupled model
(eqs.~(\ref{Nsolution},\ref{simpleI})).  Inset A shows a detail of the
temporal evolution around $t_{th}$ to see how the approximate and the full
solution deviate from each other.  Inset B shows three selected wing modes.
Values for the intensity peaks and corresponding times are given in
Table~\ref{max-time}.  Notice that in the main figure, as well as in inset A,
curves (a) and (b) are practically indistinguishable.  The arrow underneath
the label in these two figures shows the order in which the labels have to be
attributed to the (closely) adjacent curves.  
}
\label{comp-centremod} 
\end{figure}

Fig.~\ref{comp-centremod} shows the temporal evolution of a selection of modes
(those close to line center) for the full integration (eqs.~(\ref{model}),
solid lines) and for the decoupled system
(eqs.~(\ref{Nsolution},\ref{simpleI}), dashed lines).  We notice that over the
whole range of validity ($t \le t_{th}$) of the analytical solution for the
carrier number,
eq.~(\ref{Nsolution}), the decoupled approximation is
very good.  Inset A shows a detail of the temporal evolution for the same
modes close to threshold:  all curves lie extremely close to one another,
confirming the validity of our approximation.  Inset B shows the evolution for
three wing modes (cf. caption):  the agreement between full solutions (solid
lines) and analytical approximations (dashed lines) remains excellent even
beyond threshold for these modes.

\begin{figure}[ht!]  
\includegraphics[width=0.9\linewidth,clip=true]{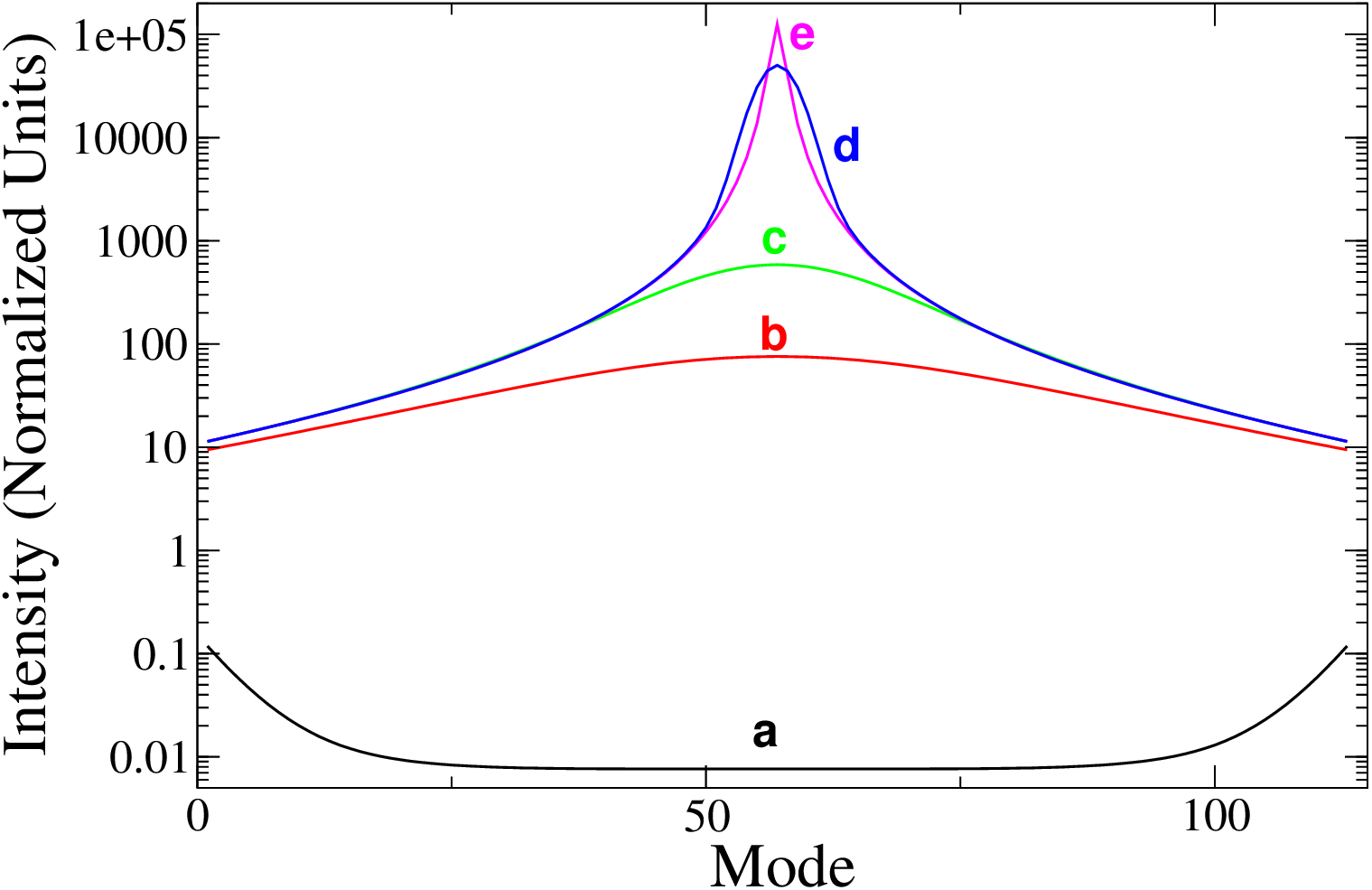}
\caption{
Intensity distribution for all modes considered in the integration, each curve
corresponding to a different time:  (a) initial condition; (b) $t = 0.62 ns$
(threshold crossing); (c) $t = 0.726 ns$ (total intensity peak); (d) $t = 1.5
ns$ (total intensity and carrier density relaxed to their asymptotic value):
beginning of the slow transient~\cite{slowmodes}; (e) $t = 35 ns$ (asymptotic
intensity distribution).  The intensity values are plotted in logarithmic
scale and are not rescaled according to the factors specified in
Fig.~\ref{time-trace}.  } \label{int-distr} \end{figure}

We summarize the main result of this section in Fig.~\ref{int-distr} which
shows how the intensity distribution among all modes at the end of the
decoupled regime (i.e., the instant when the population carrier crosses its
asymptotic value -- threshold -- for the first time) does not differ very
substantially from the initial distribution.  Curve a (black online) shows the
equilibrium intensity distribution at the initial condition -- the side modes
carry more intensity thanks to their reduced absorption (below transparency).
Curve b (red online) shows the intensity distribution at threshold (cf. above)
-- i.e., $t \approx 0.62 ns$.  Even though the central modes carry a larger
amount of intensity (and the shape of the curve is inverted, compared to the
initial condition, when comparing to the asymptotic, equilibrium, distribution
(curve d, blue online), the quantitative differences among modes at the first
threshold crossing are nearly negligible.  Anticipating on the next section,
we remark that even at the peak in the total intensity the intensity
distribution among modes (curve c, green online) differs less from the
threshold condition than from the asymptotic one.  

In conclusion, in this section we have shown that the decoupled model gives a
very good representation of the evolution of the modal content of the laser
output in the below threshold region, i.e., over a large part of
the
transient.  We have also seen, both numerically and analytically, how the side
modes carry an amount of energy much larger than their asymptotic share.
Once the system overcomes this threshold, the analytic
approximation for the carrier number breaks down, the coupling becomes strong
and only the full numerical solution can account for the remainder of the fast
evolution towards the steady state for the two global variables ($N$, and
$I_t$~\cite{slowmodes}).

\subsection{Strongly coupled regime}\label{strong-coupling}
 
The two preceding sections have analyzed in detail the portion of the
dynamical evolution below the threshold value for the carrier number.  Here we
are going to investigate the next portion of the
dynamics~\cite{def-linear-regime}, which corresponds to the full nonlinear
coupling between modes and which leads, eventually, into the time-independent
total intensity with the residual slow evolution of the modal intensity
distribution~\cite{slowmodes}.  Our analysis of the multimode high-intensity
regime is based on a direct integration of the model equations and on the
potential for the interpretation of the resulting observations which comes
directly from the detailed analysis of the previous sections.  A
direct-solution approach, such as the one adopted in~\cite{Demokan1984} for an
analytical construction of the solution for short pulses in a single-mode
device, would be very hard to adopt in our strongly multimode model.

\begin{figure}[ht!]
\includegraphics[width=0.9\linewidth,clip=true]{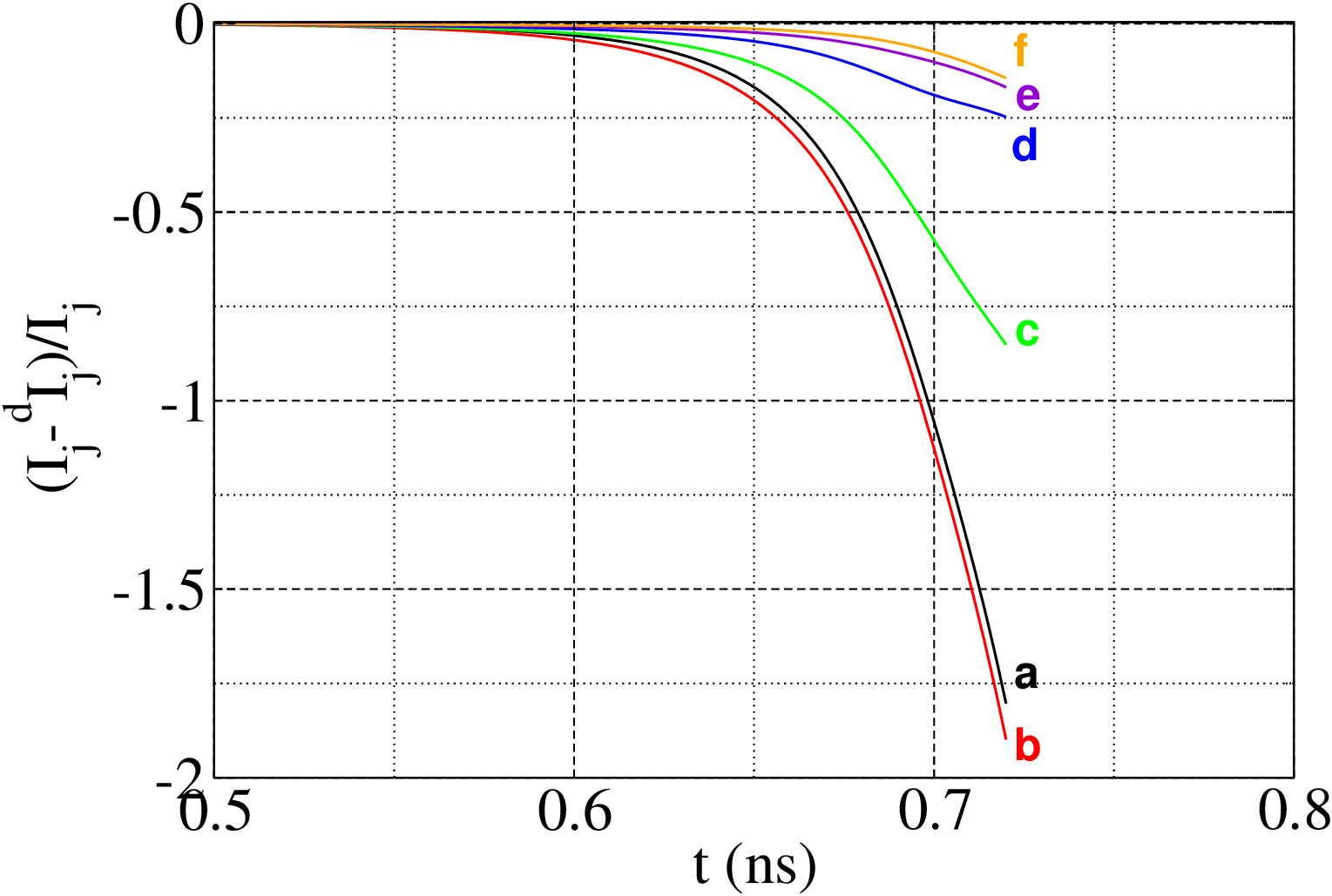}
\caption{
Relative deviation between the full-dynamics and the decoupled dynamics
$\left( \frac{I_j - ^dI_j}{I_j} \right)$ and onset of the strongly-coupled
regime.  (a) total intensity; (b) $j = 57$; (c) $j = 40$; (d) $j =
30$; (e) $j = 20$; (f) $j = 10$;
} 
\label{error-I} 
\end{figure}

Fig.~\ref{error-I} clearly shows the transition between the decoupled regime,
where the deviation between the approximate and exact modal intensities is at
most 5\% (at $t \approx 0.6 ns$), and the fully nonlinear regime, where this
difference rapidly diverges.  The transition is clearly illustrated by
Fig.~\ref{gamma-j} which shows that during their growth some $\gamma_j$'s
become positive for a subensemble of modes (near line center) and oscillate
together with the carrier number to relax~\cite{slowmodes} to their
(negative~\cite{asympt-gamma}) asymptotic
values (not shown).  When some $\gamma_j
> 0$ the nature of eq.~(\ref{simpler-approx-form}) changes, since the growth
of the $^dI_j$'s is no longer exclusively due to the growth of the prefactor
but also to the exponential growth of (some) $^dI_j$'s, which eventually leads
to a divergence from the solution of the full model, eqs.~(\ref{model}).

From the point of view of the solution of  eqs.~(\ref{model}) the positivity
of some $\gamma_j$'s bears no weight.  Indeed, throughout the growth of the
modal intensities ($t \le 0.72 ns$, cf.  Fig.~\ref{time-trace}), the r.h.s. of
eq.~(\ref{modela}) (or, equivalently, eq.~(\ref{decoupledI})) remains positive
due to the contribution of the spontaneous emission ($\mathcal{B}_j$),
independently on the sign of $\gamma_j$.  Thus, the latter contributes
quantitatively, but not qualitatively, to the growth of the modal intensities
$^dI_j$.

\begin{table}
\caption{
Intensity value at the first$(n=1)$ and second $(n=2)$ peak, $^{pn}I_j$, and
time of occurrence, $^{pn}t_j$, $^{p1,p2}\Delta t_j \equiv {^{p2}t_j}
-{^{p1}t_j}$ time difference between the peaks, and asymptotic intensity
 $\overline{I}_j$ for
each laser mode $j$ (from the full set of equations,~(\ref{model})).  The last
line, separated from the rest of the table, gives the equivalent information
for the total intensity.
}
\label{max-time}
\begin{tabular}{||c|c|c|c|c|c|c||}\hline\hline
Mode  & $^{p1}I_j$ & $^{p1}t_j(ns)$ & $^{p2}I_j$ & $^{p2}t_j(ns)$ &${^{p1,p2}\Delta t_j} (ns)$& $\overline{I}_j$ \\ \hline\hline
57 & 48182 & .7140 & 39449 & .9264 & .2124 & 125645 \\ \hline 
56 & 47111 & .7138 & 37299 & .9248 & .2110 &  41210 \\ \hline
55 & 44054 & .7134 & 31598 & .9210 & .2076 &  13627 \\ \hline
54 & 39443 & .7126 & 24136 & .9152 & .2026 &   6437 \\ \hline
52 & 27972 & .7102 & 10805 & .9010 & .1908 &   2387 \\ \hline
50 & 17190 & .7070 &  3856 & .8858 & .1788 &   1223 \\ \hline
40 &   717 & .6854 &   229 & .8786 & .1932 &    198 \\ \hline
30 &   112 & .6726 &    76 & .8682 & .1956 &     72 \\ \hline
20 &    42 & .6700 &    35 & .8670 & .1970 &     34 \\ \hline
10 &    21 & .6718 &    18 & .8694 & .1976 &     18 \\ \hline\hline
tot & 598171 & .7100 & 317103 & .9147 & .2047 & 279118 \\ \hline\hline
\end{tabular}
\end{table}

A closer look at the modal dynamics (solid lines in Fig.~\ref{comp-centremod})
shows that all individual modes display the same generic behaviour (peak with
damped oscillations) but that some differences appear in the time at which the
peaks occur and in the shape (frequency and damping) of the oscillations.
Table~\ref{max-time} gives the time $^{p_1}t_j$ at which the intensity maximum
($^{p_1}I_j$) is reached for a selection of modes.  A clear anticipation in
the first peak appears as the mode index moves away from line center ($j=57$),
not unexpectedly after the discussion of the previous section based on the
behaviour of the $\gamma_j$'s (Fig.~\ref{gamma-j}).  An inversion in this
tendency, however, appears comparing $^{p_1}t_{20}$ and $^{p_1}t_{10}$,
hinting to a more complex behaviour in the strong-coupling regime.

\begin{figure}[ht!]
\includegraphics[width=0.9\linewidth,clip=true]{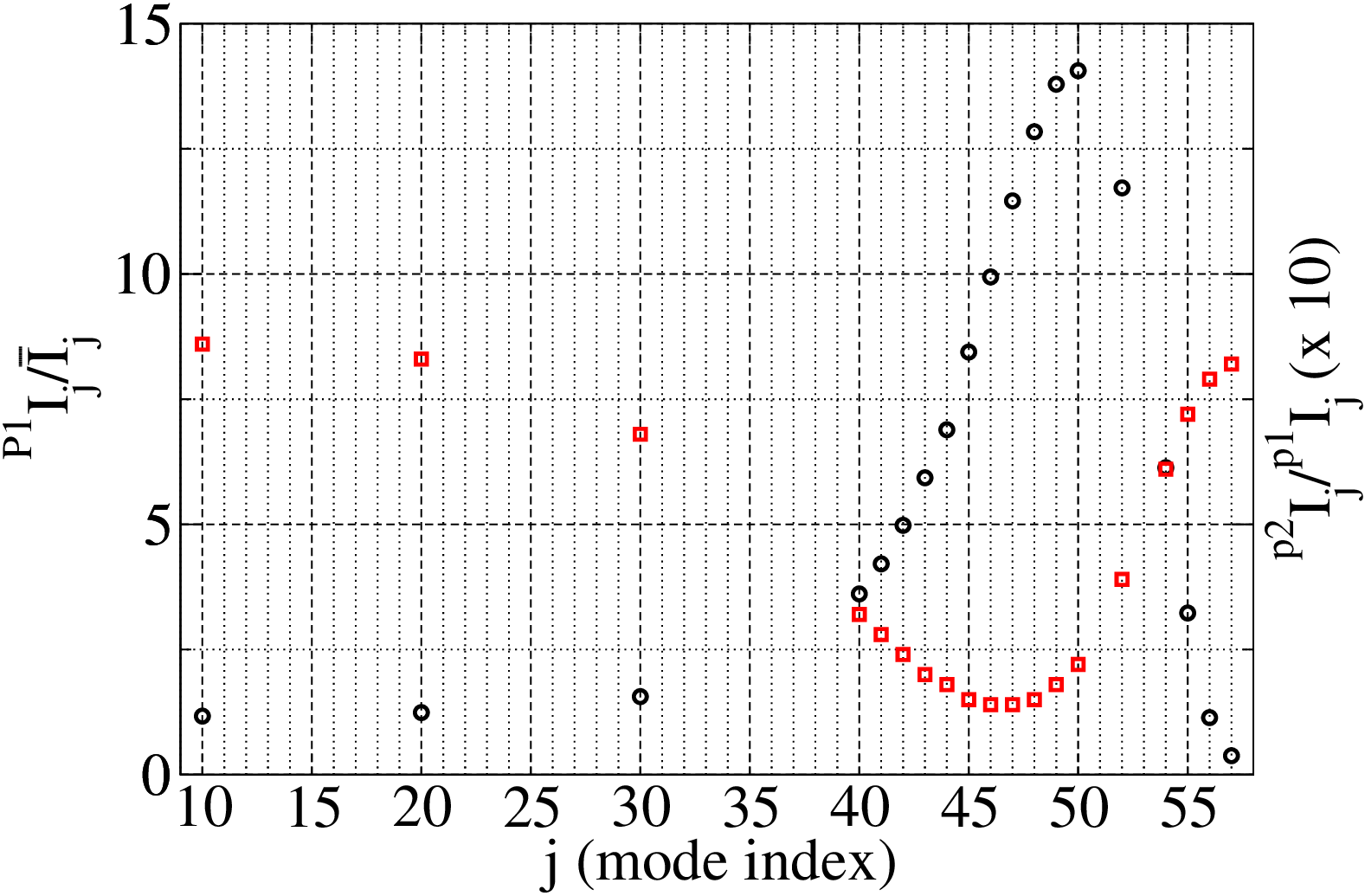}
\caption{
Modal peak amplitude relative to the modal steady state,
$\frac{^{p_1}I_j}{\overline{I}_j}$ (circles, left scale) and amplitude of the
second peak relative to the first peak, $\frac{^{p_2}I_j}{^{p_1}I_j}$
(squares, right scale) for a selection of laser modes.  Notice that the
central mode is at the far right, thus for picturing the full set of modes one
should mirror-image the figure with respect to vertical axis passing through
$j=57$.  For the total intensity, the equivalent quantities are (from
Table~\ref{max-time}):  $\frac{^{p1}I_{tot}}{\overline{I}_{tot}} = 2.14$,
$\frac{^{p2}I_{tot}}{^{p1}I_{tot}} = 0.53$.  
} 
\label{damping} 
\end{figure}

Fig.~\ref{damping} shows the ratio between the maximum peak amplitude for each
mode
and its corresponding asymptotic value (circles, left scale).  For the central
mode  the peak amplitude $^{p1}I_{57}$ is about half its steady state,
$\overline{I}_{57}$, while in the far wings the peak (e.g., $^{p1}I_{10}$) is
only slightly higher than its steady state ($\overline{I}_{10}$), showing that
equilibrium for these modes is reached rather smoothly (i.e., with little
overshoot and oscillations).  However, the peak amplitude grows very large for
modes off line-center, but not too far from it, the largest ratio occurring
for $j=50$ (six places away from line center) whose peak intensity
$^{p_1}I_{50} \approx 14 \times \overline{I}_{50}$.  This highlights the role
of the (close) side modes in the strongly coupled regime, which temporarily
carry a large amount of energy.

Associated with this strong variation in the peak intensity, a remarkable,
nearly anti-correlated, variation in the damping appears.  The ratio between
the second and the first peak $\left( \frac{^{p_2}I_j}{^{p_1}I_j} \right)$ is
represented by squares (Fig.~\ref{damping}, right scale)).  While at line
center and far off in the wings the second peak is almost as large as the
first ($\gtrsim 80\%$), for the strongest overshoot we notice (almost) the
largest damping (with a slight shift, $\frac{^{p_2}I_{47}}{^{p_1}I_{47}}
\approx 0.2$).

\begin{figure}[ht!]
\includegraphics[width=0.9\linewidth,clip=true]{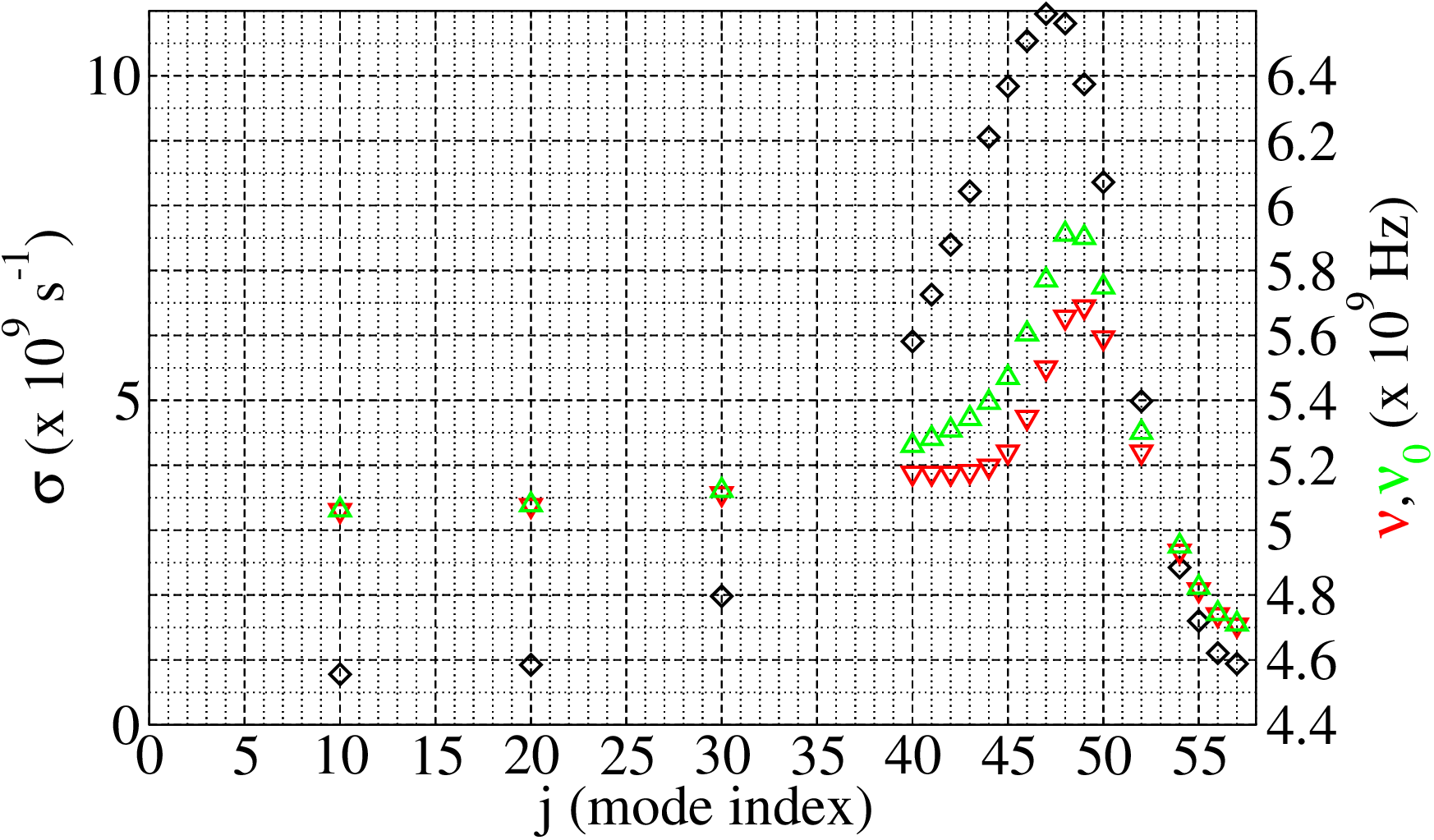}
\caption{
Modal damping coefficient, $\sigma_j$ (diamonds, left scale), modal damped
(down-triangles), $\nu_j$, and undamped (up-triangles), $\nu_{0,j}$,
relaxation oscillation frequency (right scale).  Notice that the central mode
is at the far right, thus for picturing the full set of modes one should
mirror-image the figure with respect to vertical axis passing through $j=57$.
The equivalent quantities for the total intensity are:  $\sigma_{tot} = 3.10
\times 10^9 s^{-1}$, $\nu_{tot} = 4.885 GHz$, $\nu_{0,tot} = 4.910 GHz$.
} 
\label{freqs} 
\end{figure}

Assuming a simple dependence for the damped oscillations, of the form $I_j(t)
= I_{0,j} e^{-(\sigma_j + i 2 \pi \nu_j) t}$, we can estimate the modal damping
constant $\sigma_j$ in the strongly coupled regime, together with the
oscillation frequency $\nu_j$ directly from the data, using the peak
amplitudes ($^{p_1}I_j$, $^{p_2}I_j$) and times   ($^{p_1,p_2}\Delta t_j
\equiv {^{p_2}t_j} - {^{p_1}t_j}$) at which they occur.  With the help of the
standard expression for the undamped frequency $\nu_0$ of a damped oscillator,
we can further estimate $\nu_{0,j} = \sqrt{\left( \frac{\sigma_j}{2 \pi}
\right)^2 + \nu_j^2}$.  The corresponding results are shown in
Fig.~\ref{freqs} for the damping $\sigma_j$ (diamonds, left scale) and for the
damped (down-triangles), $\nu_j$, and undamped (up-triangles), $\nu_{0,j}$,
frequencies (right scale for both).

A resonance-like behaviour appears clearly in the relaxation oscillation
frequency, with a maximum at mode $j=48$ for $\nu_0$ (at $j=49$ for $\nu$),
and a sizeable variation in frequency $\Delta \nu_0 = \nu_{48} - \nu_{57}
\approx 1.2 GHz$ (i.e., $\frac{\Delta \nu_0}{\nu_{57}} \approx 25\%$).  The
contribution due to the modal damping $\sigma_j$ is clearly visible in the
graph and increases the frequency, at its maximum, by
$\frac{\nu_{0,48}-\nu_{48}}{\nu_{48}} \approx 4 \%$, while shifting it away
from line center (by one mode).  The combination of the contribution of all
modes lends the total intensity (Fig.~\ref{time-trace}) features (cf. captions
of Figs.~\ref{damping},~\ref{freqs}) which are a combination of those of
different modes.  Looking at the damping, $\sigma_{tot}$ or at the relative
height of the first peak, $\frac{^{p1}I_{tot}}{\overline{I}_{tot}}$, the
values are similar to those of mode $j = 53$, the relaxation oscillation
frequencies, $\nu_{tot}, \nu_{0,tot}$, are consistent with mode $j=54$, while
the second peak is less strongly damped, as for a mode placed between $j=55$
and $j=56$.

We therefore conclude that the dynamics in the portion of the fast transient
corresponding to the strongly-coupled regime is dominated by a group of modes
close to, but not at, line-center.

\begin{figure}[ht!]
\includegraphics[width=0.9\linewidth,clip=true]{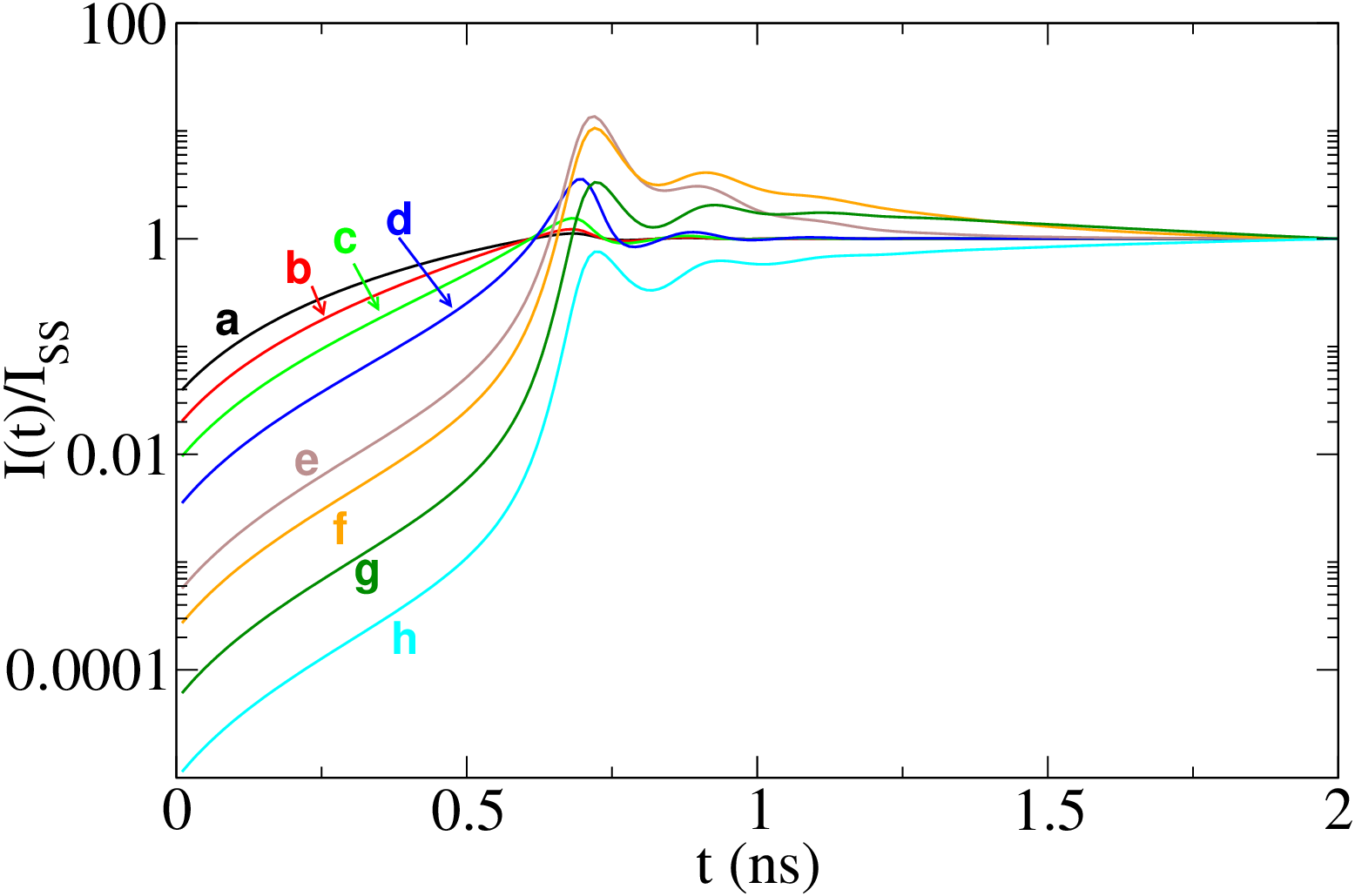}
\caption{
Temporal evolution for a sample of modes normalized to their individual
asymptotic intensity values.  To simplify the label, we have omitted the
subscript $j$ for the variable $I$ in both numerator and denominator of the
vertical axis.  (a) $ j = 10$; (b) $j = 20$; (c) $j = 30$; (d) $ j = 40$; (e)
$j = 50$; (f) $j = 52$; (g) $j = 54$; (h) $j = 57$.  Mode $j = 56$ shows the
same evolution as mode $j = 57$ (in this figure) slightly displaced towards
higher values (not shown for simplicity).  } \label{norm-peaks} \end{figure}

An equivalent graphical illustration is given by Fig.~\ref{norm-peaks} which
shows the transient evolution for each mode, normalized to its asymptotic
value.  The modes far in the wings show a very early and gradual growth with
little overshoot and oscillations (the decay is masked by the normalization to
their asymptotic state, cf.~\cite{slowmodes}).  Coming closer to line center
we notice a gradual delay in the growth, accompanied by a sharp, and very
large, peak and a gradual relaxation towards the asymptotic intensity level.
The last mode to grow is the central one, whose peak, together with the first
side mode, is lower than its steady state.  The relaxation for the central
modes is extremely slow and extends well beyond the figure
range~\cite{slowmodes}.

Fig.~\ref{norm-peaks}, together with the analysis of the previous sections,
convincingly demonstrates the reason for the slow dynamics discussed
in~\cite{slowmodes}.  The transient evolution from the initial,
below-threshold condition takes place first through a regime where the modal
growth is not strongly differentiated between mode center and wings (i.e.,
below threshold).  When the laser finally passes threshold, the central
(slower) modes have to grow away from an intensity distribution which is far
from the asymptotic one (Fig.~\ref{int-distr}).  This intensity configuration
favours those lateral modes which are sufficiently close to line center to
have a strong growth, but which at the same time possess faster time
constants, which allow them to more rapidly exploit the excess population
inversion.  Thus, we find a strongly out-of-equilibrium intensity distribution
at the peak (curve c in Fig.~\ref{int-distr}), and it is not surprising that
when the total intensity and the carrier number have relaxed to their
respective asymptotic values (at $t \approx 1.5 ns$ for the initial conditions
chosen in this paper, Fig.~\ref{time-trace}) the modal repartition of the
intensity be still strongly very far from its asymptotic configuration.  In
this way, we clearly understand the reason for the slow dynamics reported
in~\cite{slowmodes}.

\section{Transient spectra}\label{spectra}

The analytical and numerical treatment of the transient has given us a good
description of the modal evolution from the initial state (laser -- nearly --
off) to the point where the two global variables ($N$ and $I_t$) have attained
equilibrium.  We now look at the time-resolved transient emission spectrum,
defined as the frequency interval over which the modal intensity passes a
chosen percentage of the total output intensity, $W^{FS}_x$, where $FS$ stands
for Full Spectral and $W$ for Width, while $x$ fixes the chosen level in dB.
We remark that, contrary to what found for the slow dynamics~\cite{slowmodes},
the transient spectra are only marginally affected by the modal arrangement
under the gain line (even vs. odd number of modes in the simulation --
cf.~\cite{slowmodes} for a discussion).  We will therefore discuss only the
modal placement used so far.  

\begin{figure}[ht!]
\includegraphics[width=0.9\linewidth,clip=true]{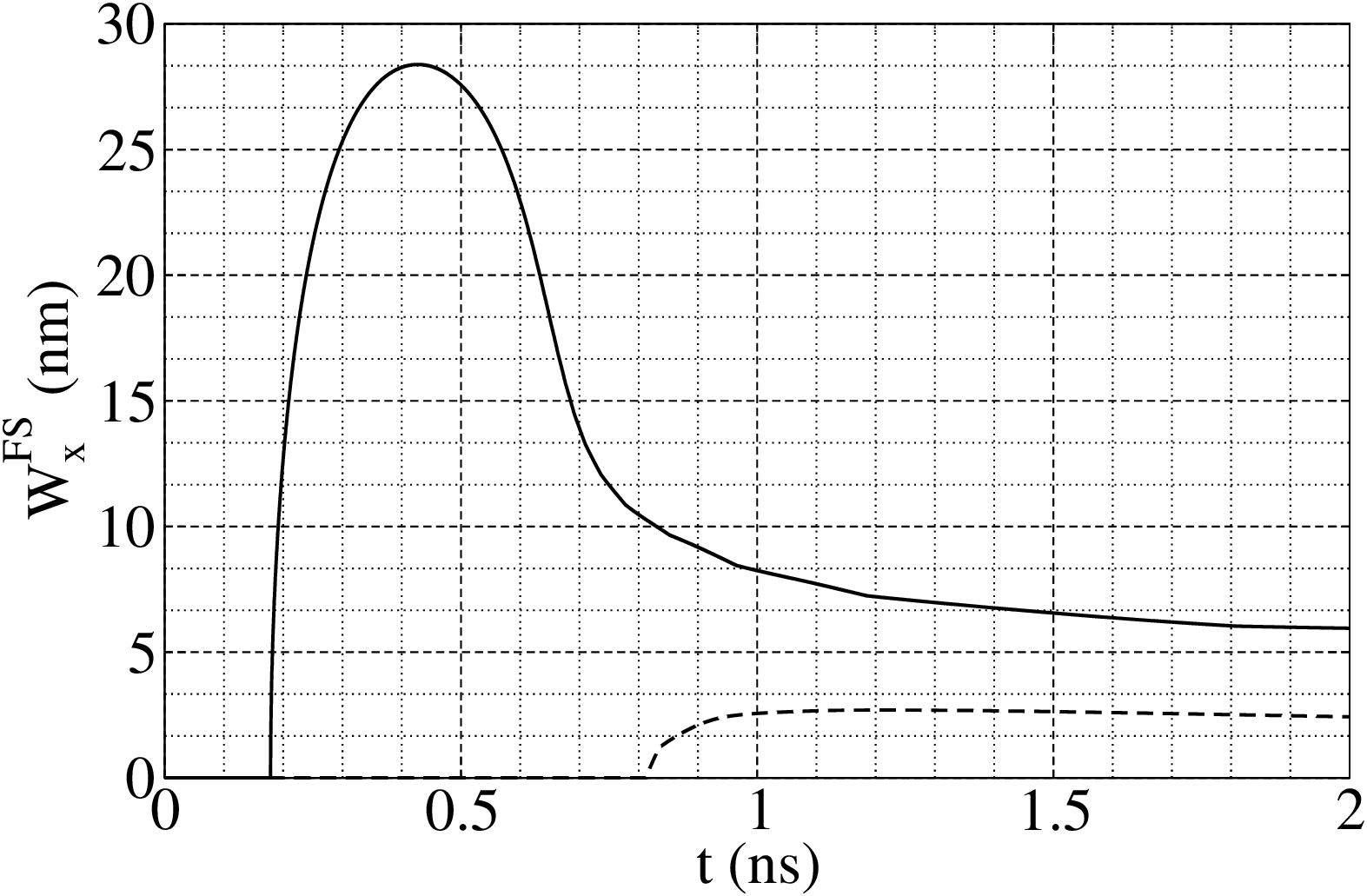}
\caption{
Full spectral width $W^{FS}_x$ of the laser evolution during the transient.
Solid lines $x = 40 dB$, dashed lines $x = 20 dB$.  The level in $dB$ is
defined, as in~\cite{slowmodes}, as the modal intensity which passes the
preset level compared to the instantaneous total intensity.  }
\label{all-spectra} \end{figure}

Fig.~\ref{all-spectra} shows the Full Spectral Width ($W^{FS}$), resolved in
time, calculated at -40 dB ($W^{FS}_{40}$, solid line) and at -20 dB
($W^{FS}_{20}$, dashed line).  The frequency ``edge", marked by each line in
the figure, is obtained by linear interpolation between the frequencies of the
two modes between which the $W^{FS}_x$ criterion is satisfied.  The -40 dB
level captures the full evolution of the spectral content during the
transient; we recognize that in the very first phases of the transient ($ <
0.2 ns$) none of the modes passes this mark (the spectral width starts at this
point)~\cite{farbelow} because the intensity is widely distributed over a
large number of modes, rather than being carried by a few modes around line
center.  In other words, no frequency component is sufficiently strong to
reach one-hundredth of the total intensity (equivalent to -40 dB).  Since we
are integrating over slightly more than 100 modes, this implies that the
intensity distribution, in this first phase of the transient, is sufficiently
homogeneous that no spectral feature emerges from the spectrum.  This is in
agreement with our previous findings which show a rather homogeneous intensity
distribution among modes in the decoupled regime.  The maximum width is
obtained between $0.4 ns$ and $0.5 ns$ and corresponds to approximately 47
modes passing the $0.01 \times I_t(t)$ level (i.e., the -40 dB mark).

The $W^{FS}_{40}$ continues to decrease beyond $t = t_p$ and $W^{FS}_{20}$
makes its appearance at $t \approx 0.8 ns$, grows until $t \approx 1 ns$ and
then initiates its slow convergence to its asymptotic value, as discussed
in~\cite{slowmodes}.  In light of the results of the previous section the wide
spectral width ($W^{FS}_{40}$) does not come as a surprise and quantitatively
shows the degree of modal spread of the laser intensity in the initial phases
of the transient.  On the other hand, the large modal coverage has a direct
impact on the validity of models which are obtained by truncating the full
one, $M = 113$, to a subset of modes around line center~\cite{otherpap}.

\section{Comments, Summary and Conclusions}\label{conclusions}

The results of this paper have been obtained on a semiconductor laser model
where specific choices have been made to describe relaxation processes to
match a specific experimental device (a Trench Buried Heterostructure bulk
semiconductor laser).  We have selected this model to have a tested, realistic
comparison with an actual device~\cite{Byrne}, but the relevance of the
understanding that we have gained in the
process extends beyond the specific details of this particular system.

Numerical tests have shown~\cite{Dokhane1990} that only
minor quantitative details change when replacing the fixed gain line with a
variable, dynamic one~\cite{Dutta1980}, or when substituting the parabolic
profile
(eq.~\ref{coeff2}) with a Gaussian one~\cite{Marcuse1983}, or even when
exchanging the linear gain (eq.~\ref{coeff2}) with a logarithmic
one~\cite{Makino1996}.  More sophisticated modeling choices can be made for
the gain line~\cite{Balle1995,Balle1998}, but on the basis of the previous
remarks, we expect the general results we have obtained in this paper to hold at
least for most kinds of multi-longitudinal mode semiconductor lasers.

The kind of interaction among modes (coupling to a mean field) is at the origin
of our numerical observations and of the understanding we have gained,
independently of the modeling details.  Thus, we expect our predictions to apply
to any other kind of laser where an intensity-based intermodal interaction
dominates.  Candidates for such behaviour are all those lasers where diffusion
in the gain medium washes out any
trace of spatial modulation resulting from the interference among the modal
fields and/or lasers with a partial, but not dominant, inhomogeneous
broadening.  

Our modeling of the below threshold region (decoupled regime,
Sections~\ref{carrier-density},~\ref{decoupled-modes}) may deliver useful
physical information on the dynamics of carriers and optical emission in cavity
light emitting diodes (CLEDs).  Two particular subgroups of CLEDs may benefit
from this modeling: superluminescent diodes, which are very sensitive to
parasitic cavity effects and may suffer from parasitic lasing, and edge emitting
LEDs (ELEDs), which are the LED equivalent of an edge-emitting semiconductor
laser, exactly the kind of device which is the object of the model we have
investigated~\cite{Byrne}.

Finally, we recall what stated in the introduction about the generalization of
the model to a set of oscillators, possessing a stable attractor and mutually
coupled through a mean field. The dynamics of such systems, including the
presence of a master mode dominating the slow evolution, will qualitatively
match the one described in our paper, when externally driven, irrespective of
the details of the model.

In summary, we have numerically investigated the fast transient dynamics of a
multimode semiconductor laser, in response to a sudden turn-on by a switch of
a parameter (pump), modeled by M ODEs for the modal intensities, with coupling
occurring through the population inversion (i.e., the carrier number).  The
dynamics have been proven to separate into two regimes:  (a) one where the
carrier number and the modal intensities can be decoupled and for which
approximate analytical solutions can be found; and (b) the strongly coupled
(or fully nonlinear) one where all variables are interdependent in their
evolution.  Our analysis gives a clear physical explanation and a quantitative
illustration of the origin of the strong deviation for the modal intensity
distribution at the onset of equilibrium ($t = \overline{t}$) for the global
variables (carrier number $N$ and total intensity $I_{tot}$) which gives rise
to the slow dynamics~\cite{slowmodes}.


\begin{thebibliography}{}
\bibitem{initialbias} Choosing a very low initial bias allows us to
investigate the various regions of the transient.  Practically useful bias
values, such as those of ~\cite{slowmodes}, will
skip the transient evolution below transparency, thus shortening and
simplifying the dynamics.
\bibitem{alt-def-N} The choice of using $\mu_2$ is entirely arbitrary and
eqs.~(\ref{expl-an-sol},\ref{expl-an-coeff}) can be written in terms of
$\mu_3$ as well, since in the polynomial terms only $\mu_2$'s modulus or real
part enter into the equation, while in the exponential form, a simple sign
change restores the form with $\mu_3$ in place of $\mu_2$.
\bibitem{check-an-num} We have checked that the numerical solution obtained
from this decoupled set of equations coincides with the analytical one within
the numerical precision, for the same initial condition (in practice
deviations of a few parts per thousand are tolerable).  The advantage of using
the numerical result comes from the synchronicity of the points for the two
traces:  computing both with the help of the set of ODEs provides synchronous
ensembles of points, while those originating from 
eq.~(\ref{Nsolution}) are given at times $t$ which cannot be synchronized
(since $t$ is the dependent variable in this expression).
\bibitem{diff-gamma-beta} One should define two different kinds of coefficients
for $\gamma$ and $\mathcal{B}$, depending on whether they are defined as a
function of $N$ or of $\mathcal{N}$.  Due to the practically identical results
(cf. Figs.~\ref{gamma-j},~\ref{beta-j}), originating from the almost
coincidence of $N$ and $\mathcal{N}$ over the whole interval of
validity of the decoupled approximation (cf. Fig.~\ref{sols-N}), for the sake of
simplicity we refrain from introducing additional symbols.
\bibitem{transp} The point at which all curves for $\gamma_j$ cross
corresponds to $N=N_0$, which represents the carrier number for which the gain
medium is bleached (transparency).  Below this value, the wing modes
experience less absorption, since they are far off-resonance, thus their
corresponding $\gamma_j$'s are larger.  For $N>N_0$ the situation is reversed,
since the gain is larger near line center.
\bibitem{constr-bth} Recall that we consider the time interval $(0 \le t
\lessapprox 0.6) ns$ (cf. Fig.~\ref{t113-sum}), thus this statement does not
cover the strongly coupled regime ($t \gtrapprox 0.62 ns$) where the modal
intensities show a peak.
\bibitem{winggain} The larger intensity increment below transparency is
explained, for the wing modes, by their lower absorption.  Notice that for the
given lineshape, eq.~(\ref{coeff2}), for a couple of modes at the extrema of
the interval there may be a (temporary) sign inversion, which bears, however,
no practical consequences on the numerics.
\bibitem{convergence-radius} Since the radius of convergence of the exponential
function is infinite, one could use a less restrictive condition.  However, the
magnitude of $\Delta t$ is appropriately set this way and is consistent with
the one used in Fig.~\ref{ratio-int}.
\bibitem{def-linear-regime} An alternative way of defining the upper limit for
the range of parameters (or time interval) for which the dynamical decoupling
between carrier number and modal intensities holds is the following:  evaluate
steady states for the variables $N$ and $I_j$ in the decoupled model
(eqs.~(\ref{Nsolution},\ref{simpleI}) and impose the condition that $I_j \ge
0$ $\forall j$.
While well below transparency three distinct real solutions exist for the
steady state condition (eq.~(3b)
in~\cite{slowmodes}, adapted to the decoupled system), close to threshold only
one real one, which we will call $\mathbb{N}$, can be found.  Thus, the steady
state 
intensity takes the form (for all modes) $ \overline{I}_j = \frac{B_j
\mathbb{N}\left( \mathbb{N}+P_0 \right)}{K_j - \alpha_j \mathbb{N}}$, thus
$\overline{I}_j > 0$ iff $K_j - \alpha_j \mathbb{N} = - \gamma_j  >  0.$
Hence, the decoupled regime holds only for $\gamma_j < 0, \forall j$ and the
instant $\hat{t}$ for which $\gamma_j(\hat{t}) > 0$ (in particular $j=57$)
defines the upper limit to this regime (either by defining $\hat{t}$ or the
corresponding value $\mathcal{N}(\hat{t})$).\\ Thus, for the decoupled
regime to hold, the effective relaxation constant must remain negative at all
times for all modes.  As soon as one $\gamma_j > 0$ ($j=57$, specifically) the
approximation can no longer be used.
\bibitem{asympt-gamma} The asymptotic negativity of $\gamma_j$ is ensured by
the fact that $\left. \frac{dI_j}{dt} \right|_{t \rightarrow \infty} = 0$
while $\mathcal{B}_j > 0$, $\forall j$.
\bibitem{farbelow} The higher the laser prebias, the earlier the $W^{FS}_{40}$
spectrum starts.  Here, the initial condition is the one used throughout the
paper (injected current $J_i = 1 mA$).
\bibitem{otherpap} This point will be discussed in detail elsewhere.
\end{thebibliography}
\end{document}